\documentclass[fleqn,usenatbib]{mnras}

\usepackage{amsmath}    

\IfFileExists{newtxtext.sty}{%
  \usepackage{newtxtext,newtxmath}%
}{%
  \usepackage{txfonts}%
}
\usepackage[T1]{fontenc}
\usepackage{changes}

\DeclareRobustCommand{\VAN}[3]{#2}
\let\VANthebibliography\thebibliography
\def\thebibliography{\DeclareRobustCommand{\VAN}[3]{##3}\VANthebibliography}

\usepackage{orcidlink}
\usepackage{graphicx}   
\usepackage{booktabs}   
\usepackage{tabularx}   
\usepackage{xcolor}     

\providecommand{\nodata}{\ensuremath{\cdots}}

\usepackage{etoolbox}
\robustify\cite
\robustify\citep
\robustify\citet
\robustify\citealt
\robustify\citeauthor
\robustify\citeyear

\title[Variably irradiated tidally locked planets]{Atmospheric dynamics and variability of periodically
irradiated tidally locked planets}

\author[D. Banik et al.]{
Deepayan Banik\orcidlink{0000-0002-7179-8254},$^{1}$
Vasilii Pustovoit\orcidlink{0009-0007-3199-8511},$^{1}$
Thaddeus D. Komacek\orcidlink{0000-0002-9258-5311}$^{2}$\thanks{E-mail: tad.komacek@physics.ox.ac.uk}
and Marta L. Bryan\orcidlink{0000-0002-6076-5967}$^{3}$
\\
$^{1}$Department of Physics, University of Toronto, 60 St George Street, Toronto, ON M5S 1A7, Canada\\
$^{2}$Atmospheric, Oceanic and Planetary Physics, Department of Physics, University of Oxford,
Parks Road, Oxford OX1 3PU, UK\\
$^{3}$Department of Astronomy and Astrophysics, The Pennsylvania State University,
525 Davey Laboratory, University Park, PA 16802, USA
}

\date{Accepted XXX. Received YYY; in original form ZZZ}
\pubyear{2026}

\begin{document}
\label{firstpage}
\pagerange{\pageref{firstpage}--\pageref{lastpage}}
\maketitle

\begin{abstract}
Time-variable stellar heating may create new climate states, such as permanently hot or cold hemispheres, on asynchronous planets. Planets on eccentric orbits, circumbinary planets and planets on polar orbits around gravity-darkened stars all receive variable irradiation, yet variably irradiated planets (VIPs) have not been studied from a fundamental dynamical perspective. We present the first generalized study of the atmospheric response of variably irradiated \textit{tidally locked} planets, using a 1.5-layer shallow-water model forced towards a time-dependent radiative equilibrium. We derive linearized analytic expressions for the mean, amplitude and phase lag of the response, which is that of a forced damped harmonic oscillator, and recover the established trends of tidally locked planets in the limit of steady forcing. When the instellation variation and wave timescales match and the damping is weak, a resonance occurs because waves propagate and interfere freely. The resonance produces a global standing wave mode and consequently a temporarily warmer nightside. Linear and nonlinear simulations reproduce these results and show that the hotspot oscillates longitudinally at the forcing frequency and reaches the antistellar point at resonance. The oscillation is eastward or westward of the substellar longitude depending on the rotation rate and the nonlinearity of the system. Synthetic phase curves carry signatures of this variability and invert in shape at resonance because of nightside warming. Among observed systems, we find 11 planets with at least 10\% phase curve modulation from variable irradiation. Eccentric, pseudo-synchronous giants are the most promising targets, with HAT-P-2b having the largest predicted modulation.
\end{abstract}

\section{Introduction}
\label{sec:intro}

Many close-in exoplanets are \textit{tidally locked}, that is they have a permanent dayside and nightside \citep{showman2002atmospheric,PierrehumbertHammond2019}. The
resulting day--night temperature contrast is set by a competition between radiative relaxation and advective and wave-driven transport. Equatorial waves forced by the zonally asymmetric heating propagate across the planet to move heat to the nightside, which reduces the contrast. Under strong irradiation or drag the contrast is maximized \citep{perez2013atmospheric,komacek2016atmospheric}. The
same wave adjustment also drives equatorial superrotation and an eastward-shifted hotspot \citep{ShowmanPolvani2011,hammond2021rotational}, and has been shown to reproduce observed brightness temperature contrasts across hot Jupiters \citep{komacek2017atmospheric} and modelled tidally locked rocky planets \citep{koll2016temperature,koll2022scaling}. In all of this work the instellation is assumed to be constant in time.

However, a tidally locked planet may receive instellation that periodically varies. For example, planets on eccentric orbits are driven towards pseudo-synchronous spin by tides \citep{Hut1981}. Their incident flux peaks and decays once per orbit, as is the case for planets like HD~80606b, HAT-P-2b and GJ~436b \citep{LangtonLaughlin2008,Lewis2013,Lanotte2014}. Another situation arises for misaligned/polar orbit planets around gravity darkened stars. These are rapidly rotating early-type stars with strong latitudinal temperature variations caused by the von-Zeipel effect \citep{vonZeipel1924}. Examples of planets orbiting these stars include KELT-9b, Kepler-13Ab, and WASP-189b \citep{Ahlers2020,Masuda2015,patel2026tess}. Additionally, there are pulsating host stars resulting in changing incident flux, as in the case of planet WASP-33b \citep{vonEssen2020}. For asynchronous planets, e.g., circumbinary planets, the relative binary gyration and eclipsing are sources of instellation variation \citep{may2016examining}. Variably irradiated planets (VIPs) are common.

VIPs have also been studied using numerical models. The periastron heating pulse on eccentric pseudo-synchronous planets has been shown to drive fast transient flows and long-lived vortices \citep{LangtonLaughlin2008,Iro2010,mendoncca2021variable}. Three-dimensional models retain a strong equatorial jet, as for constantly irradiated planets, but the peak infrared emission shifts relative to periastron, the effect showing up on phase curves, as modelled for HAT-P-2b and GJ~436b \citep{Kataria2013,Lewis2010,Lewis2014}. On the other hand General Circulation models (GCMs) of circumbinary planets have shown little climatic departure from an equivalent single star case \citep{may2016examining}. Most of these previous studies, however, use complex 3D GCMs that make a parameter space survey difficult, and hence a generalized understanding of VIPs is lacking. 

Recently, \cite{Banik2025} studied a range of variably irradiated asynchronous planets using a 1.5 layer shallow water model (SWM) similar to \cite{perez2013atmospheric} and showed that variable irradiation can produce previously unidentified climates on asynchronous planets, so that the asynchronous planets mimic synchronous atmospheres under instellation--diurnal resonance. A shallow water model is idealized, so the model can explore a wider parameter space than GCMs, and the reduced-order nature of the model gives greater intuition into dynamical mechanisms. This model has thus been widely used in the geophysical fluid dynamics literature alongside previous exoplanet studies \citep{Matsuno1966,philander1984unstable, penn2017thermal,ohno2019atmospheres}. 

In this paper, we build on the same SWM for variably irradiated tidally locked planets as in \cite{Banik2025} with the goal of providing a generalized theory for their atmospheric response. Aside from a linearized analytical treatment that produces closed-form solutions for the planetary response, our numerical setup also allows us to study emergent atmospheric variability at low cost. A specific indicator of this for tidally locked planets is the variability of the hotspot which has been examined before in several contexts. Strong toroidal magnetic fields can drive periodic east--west reversals \citep{RogersShowman2014,Hindle2019,Hindle2021a}, purely hydrodynamic models produce intrinsic variability in hotspot position and brightness \citep{Cho2003,KomacekShowman2020}, and westward offsets as well as epoch-to-epoch changes have been measured \citep{Dang2018,Armstrong2016}. The closest analogue to the phenomenon reported here is the periodic hotspot motion seen in models of eccentric hot Jupiters \citep{Kataria2013,Lewis2014}. However, the conditions under which such oscillations grow and stabilize are unknown, which is where our generalized theory comes in. We isolate global-scale hotspot oscillations that arise purely from the wave-dynamical response of a purely hydrodynamic atmosphere to a periodically time-varying instellation, without invoking magnetic or intrinsic variability effects. For analytical convenience we use a sinusoidal forcing, similar to \cite{Banik2025}, but our framework could, in principle, be extended to any form of instellation variation described previously, allowing for an extended dynamical study of such systems.

The remainder of this paper is organized as follows. In Section~\ref{sec:theory} we develop the linearized shallow water framework for variably irradiated tidally locked planets. From the full nonlinear equations, we derive scaling relations for the steady-state height contrast and characteristic wind speed, identify the conditions under which the linear approximation holds, and solve the time-dependent equations analytically for sinusoidal instellation forcing. This yields closed-form expressions for the amplitude and phase lag of the atmospheric response to variable irradiation. In Section~\ref{sec:simulations} we describe the numerical simulations used to validate the theory, covering Earth-like and gas giant parameter regimes across 200 shallow water simulations spanning both linear and nonlinear forcing amplitudes. In Section~\ref{sec:discussion} we discuss atmospheric variability in terms of hotspot oscillations, which turns out to be a function of rotation rate, and nonlinear effects. Then we construct synthetic phase curves illustrating the observable consequences of variable irradiation on planetary emission. Finally, in Section \ref{sec:summary} we summarize our findings, and probe a population of observed planets and place them on theoretically obtained parameter spaces and characterize their suitability for observational studies. 

\section{Linearized shallow atmospheres}
\label{sec:theory}

The three-dimensional structure of a planetary atmosphere often precludes a first-order understanding of fundamental dynamical mechanisms because of its inherent complexity. On the other hand, one-dimensional models do not account for processes such as wind motion and heat redistribution, which significantly impact observable parameters. A two-dimensional shallow water model of planetary atmospheres offers a computationally inexpensive yet dynamically accurate trade-off. We use the same spherical 1.5-layer shallow water model as in \cite{Banik2025} to study the atmospheres of variably irradiated tidally locked planets. The model allows for a manageable parameter space survey in addition to capturing leading-order circulation physics, and has been extensively used in previous work \citep{perez2013atmospheric,penn2017thermal, ohno2019atmospheres}. It features an active (driven) upper layer $h(\lambda, \phi, t)$ over a passive (inert) layer of infinite depth effectively mimicking the dynamics of the outer radiatively driven atmospheres of planets. The governing equations for momentum and mass balance are,
\begin{equation}
    \frac{\partial \mathbf{u}}{\partial t} + (\mathbf{u} \cdot \nabla) \mathbf{u} + f\mathbf{k} \times \mathbf{u} + g\nabla h = \textbf{R} - \frac{\mathbf{u}}{\tau_{\text{drag}}}, \label{eq:SW_mom}
\end{equation}
\begin{equation}
    \frac{\partial h}{\partial t} + \nabla \cdot (h\mathbf{u}) = \frac{h_{\text{eq}}-h}{\tau_{\text{rad}}} \equiv Q, \label{eq:SW_mass}
\end{equation}
\begin{equation}
   \text{with}\quad \textbf{R} = -\frac{\mathbf{u}}{h} \cdot\frac{Q+|Q|}{2},
\end{equation}
where $g$ is the acceleration due to gravity on the planet, $\mathbf{u}$ is horizontal velocity, $f = 2\Omega\sin\phi$ is the Coriolis parameter with rotation rate $\Omega$, $\lambda$ is longitude and $\phi$ latitude, $\tau_{\text{drag}}$ is the Rayleigh drag timescale, and $\tau_{\text{rad}}$ is the Newtonian cooling timescale. The term $\mathbf{R}$ is a nonlinear vertical exchange term that ensures conserved mass/momentum exchanges between the two layers, is required for the emergence of equatorial superrotation, and ensures independence from initial conditions alongside drag \citep{ShowmanPolvani2010, ShowmanPolvani2011, liu2013atmospheric}. Here, fluid height acts as a proxy for temperature and hence all height fields may be interpreted as temperature fields.

The time evolution equations above are important for the variable instellation problem of this work. However, we shall start by extracting information about dominant phenomena via analytical methods before solving the equations numerically. We do so in the following sections in two steps, first by a scaling analysis and second by a first-order reduction based on standard assumptions.

\subsection{Steady-state Scaling Analysis} \label{sec:steady_scaling}

We begin by studying the steady state, or the expected final state of the planetary atmosphere when instellation does not vary. Similar to \cite{banik2025meridional}, we perform a scaling analysis to obtain the dominant scales of unknown quantities such as wind velocity and day--night temperature contrast of the system in terms of known planetary parameters. For that, we linearize the system on a rotating f-plane \citep{vallis2017atmospheric}, about a basic state with zero background flow or velocity $\mathbf{u}_0 = 0$ in the frame of reference of a planet with global mean height $H$. Small perturbations are added on top of the base state i.e. $\mathbf{u} = \mathbf{u}_0 + \mathbf{u}', \quad h = H + h'$, with $|\mathbf{u}'|\ll \sqrt{gH} = c$ and $|h'|\ll H$. Here, $c$ is the wave speed of the gravest shallow water gravity mode responsible for heat (height) redistribution at first order. We also decompose $h_{\rm eq} = H + h_{\rm eq}'$, for mathematical convenience. Quadratic and higher-order terms produced from substitution of these into the main equations are dropped to get rid of nonlinear terms, including the nonlinear vertical exchange term $\textbf{R}$. Finally, we combine the Coriolis and drag terms to form the \textit{effective drag} term
\begin{equation}
    \frac{\mathbf{u}'}{\tau_{\rm drag}} + f\mathbf{k}\times \mathbf{u}' = \left(\frac{1}{\tau_{\rm drag}} + 2\Omega \sin \phi_0\right) \mathbf{u}' = \frac{\mathbf{u}'}{\tau_{\rm eff}}.
    \label{eq:tau_eff}
\end{equation}
Here $\phi_0$ is the reference latitude at which the $f$-plane is centred. $\tau_{\rm eff}$ is the effective drag timescale which is a parallel sum of the individual drag and rotational timescales. We acknowledge that rotational and drag effects can be dramatically different especially considering that they act perpendicular to one another and have different latitudinal sensitivity. However, as we show below, to first order both terms contribute similarly to inhibit global heat transport \citep{perez2013atmospheric,Banik2025}, and hence may be combined together for reduced-order analysis.


For the steady state we drop the time derivatives. We define the characteristic scales as the horizontal length $L \sim a$ (planetary radius), wind velocity $U$, day--night equilibrium height contrast of the forcing $\Delta h_{\rm eq}$, resulting day--night height contrast $\Delta h$, and nightside thickness $H$. The linearized steady equations result in the following momentum and mass balances in terms of scaling quantities,

\begin{equation}
    0 = -g \nabla h' - \frac{\mathbf{u}'}{\tau_{\rm eff}} \implies g \frac{\Delta h}{a} \sim \frac{U}{\tau_{\rm eff}},
    \label{eq:scale1}
\end{equation}

\begin{align}
   \text{and} \quad 0 = -H \nabla \cdot \mathbf{u}' + \frac{h'_{\rm eq} - h'}{\tau_{\rm rad}} \nonumber\\ \implies H \frac{U}{a} \sim \frac{\Delta h_{\rm eq} - \Delta h}{\tau_{\rm rad}}.
    \label{eq:scale2}
\end{align}
We note that the equality $h'_{\rm eq} - h'=\Delta h_{\rm eq} - \Delta h$ also follows from Section 4 of \cite{perez2013atmospheric}. 
We use the gravity wave speed $c = \sqrt{gH}$ mentioned before to define the wave timescale as $\tau_{\rm wave} = a/c$, which is effectively the planetary scale wave crossing time or the time taken for a pressure/height perturbation to travel across the planet. When we eliminate $U$ from the above equations, we obtain the following scaling law for the planetary day--night contrast,
\begin{equation}
    \Delta h \sim \frac{\Delta h_{\rm eq}}{1 + \cfrac{\tau_{\rm rad} \tau_{\rm eff}}{\tau_{\rm wave}^2}}   
    \text{ or} \hspace{2mm} \frac{\Delta h}{\Delta h_{\rm eq}} = \frac{1}{1+ 1/K},
    \label{eq:scale3}
\end{equation}
where $K = \tau_{\rm wave}^2 / (\tau_{\rm rad}\tau_{\rm eff})$ represents the nondimensional heat retention parameter. From the expression for $K$, a higher value implies slowly propagating waves (large $\tau_{\rm wave}$), a smaller cooling time $\tau_{\rm rad}$, or stronger drag or rotation (small $\tau_{\rm eff}$), all of which result in poor day--night heat transfer. This means that higher $K$ implies better dayside heat retention and a stronger temperature contrast. The wind driven temperature contrast of the planet relative to its forced equilibrium ${\Delta h}/\Delta h_{\rm eq}$ thus ranges from 0 to 1 with increasing $K$. This is commensurate with the drag and rotation dominant scaling laws for hot Jupiters derived in \cite{perez2013atmospheric}, except here we have a more condensed version combining the effects of rotation and drag in a single term, an analogue for which for a 3D atmosphere has been derived in Appendix A of \cite{zhang2017effects}. Consequently, the Froude number (Fr) of the system can be shown to be of the form
\begin{equation}
    \text{Fr} = \frac{U}{a/\tau_{\rm wave}} \sim 
    \frac{\tau_{\rm eff}}{\tau_{\rm wave}} \cdot
                      \frac{\Delta h_{\rm eq}/H}{1+1/K}.
    \label{eq:scale4}
\end{equation}
Fr determines whether the characteristic flow speed is larger or smaller than its dominant wave speed. This results in a velocity scaling that corresponds purely to thermal forcing, and yet has a similar structure to that of momentum forcing derived in \cite{komacek2025limited}. 

So far we have obtained the dominant scales for the steady state planetary response to constant irradiation, which matches with expectations from previous works. However, we note that these scalings hold only when the system response is linear. When nonlinear terms are included, the scalings may change and are no longer analytically tractable. In the following section, we demarcate where our linear approximation breaks down and hence we must resort to numerical solutions.

\subsubsection{Conditions for linearity}

The linear approximation holds only when the nonlinear terms in the momentum and continuity equations are negligible compared to the corresponding linear terms. We identify conditions corresponding to Equations \eqref{eq:SW_mom} and \eqref{eq:SW_mass}.
The height or mass linearity condition is satisfied when the linearized height advection $\nabla \cdot (h\mathbf{u}')$, which scales as $\Delta hU/a$, is small compared to the mean mass flux $H \nabla \cdot \mathbf{u}' \sim HU/a$, or simply the perturbation height being smaller than the background,
    \begin{equation}
     \varepsilon_h = \frac{\Delta h}{H} = \frac{\Delta h_{\rm eq}/H}{1+1/K} \ll 1. \label{eq:lin_condt2}
    \end{equation}
The momentum linearity condition is satisfied when the advection term $(\mathbf{u}' \cdot \nabla)\mathbf{u}'$, which scales as $U^2/a$, is small compared to that of drag $U/\tau_{\rm eff}$. When we substitute $U$ from Equation \eqref{eq:scale4}, we obtain,
    \begin{equation}
    \varepsilon_u = \frac{U^2/a}{U/\tau_{\rm eff}} = \frac{\tau_{\rm eff}^2}{\tau_{\rm wave}^2} \cdot \frac{\Delta h}{H} = \mathrm{Ro}_{\rm eff} \ll 1. \label{eq:lin_condt1}
 \end{equation}
where Ro$_{\rm eff}$ is the drag-modified Rossby number of the system. In the absence of drag, this reduces to the regular Rossby number wherein rotation dominates inertia and is the primary balance for the pressure gradient.

Both conditions are strongly dependent on the forcing strength relative to the base state height $\Delta h_{\rm eq}/H$. Hence, strongly irradiated planets like hot Jupiters are more likely to be in the nonlinear regime compared to relatively distant planets like sub-Neptunes. \cite{perez2013atmospheric}, studying hot Jupiters, chose $\Delta h_{\rm eq}/H$ to be 0.001 and 1 to span the linear and nonlinear regimes without complete justification. However, that is not sufficient to ensure either linear or nonlinear behaviour. Both \eqref{eq:lin_condt2} and \eqref{eq:lin_condt1} must be satisfied for linearity. In addition to the base relative response $\Delta h/H$ being much smaller than unity, condition \eqref{eq:lin_condt1} requires 
\begin{equation}
    \tau_{\rm eff}^2 \ll \tau_{\rm wave}^2.
    \label{eq:condt_inertial}
\end{equation}
Thus, for nonlinearities to be suppressed, the effective drag of the system must be large enough to damp the nonlinear response that originates from wave interactions. Commensurate with this, we may compare Figures 3 and 4 in \cite{perez2013atmospheric} to note that the differences between linear flows featuring off-equatorial Rossby gyres and nonlinear flows with equatorial superrotation are prominent only at low drag values where wave dissipation is ineffective. In our context, this corresponds to low $K$ values which automatically satisfies inequality \eqref{eq:lin_condt2}, but violates \eqref{eq:lin_condt1} leading to nonlinearities in the momentum balance that cause eddy momentum flux convergence, spinning up an equatorial jet \citep{ShowmanPolvani2011}. Consequently, the nonlinear effect can only be muted by lowering the strength of the forcing $\Delta h_{\rm eq}$, that then controls the linear-nonlinear boundary as in \cite{perez2013atmospheric}. Hence, momentum nonlinearity gives rise to important dynamical phenomena. On the other hand, the dynamics that result from the opposite effect, nonlinear mass and linear momentum equations, will be different. However, we find their impact to be less relevant. A further discussion on linearity conditions is included in Appendix \ref{sec:app_linearity}.

In the following section, we maintain both the inequalities to ensure linear behaviour. This enables us to reduce our equations to simpler forms for which closed-form analytical expressions of planetary response to unsteady forcing are possible. Later, when we run simulations across various planetary configurations, we relax these conditions and retain nonlinear terms, allowing us to test the limits of our analytical theory against the numerical solutions.

\subsection{Linearized unsteady equations for time varying instellation} \label{sec:second_order}

The goal of this section is to move on from scaling analysis and work with the full time-dependent equations in the linearized limit to find analytical solutions to relevant physical quantities like temperature contrast and average wind speed. Here, we are in the linearized limit and hence may drop the nonlinear terms from Equations \eqref{eq:SW_mom} and \eqref{eq:SW_mass} to simplify our equations. To capture the effect of changing irradiation on the atmospheric height, we retain the unsteady term in \emph{both} equations. This is the most general linear description of the system. The linearized momentum and mass balances become
\begin{equation}
  \frac{\partial \mathbf{u}'}{\partial t} + \frac{\mathbf{u}'}{\tau_{\rm eff}} + g\,\nabla h' = \mathbf{0},
\label{eq:lin_mom_unsteady}
\end{equation}
\begin{equation}
  \frac{\partial h'}{\partial t} + H\,\nabla\cdot\mathbf{u}'
  = \frac{\,h_{\rm eq}' - h'}{\tau_{\rm rad}}.
\label{eq:lin_cont}
\end{equation}

We now eliminate the velocity between the two equations. When we take the divergence of Equation \eqref{eq:lin_mom_unsteady}, we obtain $\left(\partial_t + \tau_{\rm eff}^{-1}\right)\left(\nabla\cdot\mathbf{u}'\right) = -g\,\nabla^2 h'$, and when we apply the same operator $\left(\partial_t + \tau_{\rm eff}^{-1}\right)$ to Equation \eqref{eq:lin_cont}, we obtain a single equation for the height,
\begin{align}
   \frac{\partial^2 h'}{\partial t^2}
   + \left(\frac{1}{\tau_{\rm rad}} + \frac{1}{\tau_{\rm eff}}\right)\frac{\partial h'}{\partial t}
   - gH\,\nabla^2 h'
   + \frac{h'}{\tau_{\rm rad}\tau_{\rm eff}} \nonumber \\
   = \frac{1}{\tau_{\rm rad}}\frac{\partial h'_{\rm eq}}{\partial t}
   + \frac{h'_{\rm eq}(\lambda,\phi,t)}{\tau_{\rm rad}\tau_{\rm eff}} .
  \label{eq:lin_wave_pde}
\end{align}
The form of this equation resembles a damped wave equation with radiative relaxation, and is a generalized second-order equation in time representing the dynamics of planetary response to any form of spatially and temporally varying radiation. 


Let us assume invariance in the meridional direction ($\phi$) and reduce the spatial dimension of the problem to one. This is a strong assumption because the Coriolis parameter $f$ is a function of latitude. However, the assumption is valid for subtropical motions in general and is able to capture leading-order physics of the system. For simplicity, we shall also move to arc-length coordinates and consider longitude being represented as distance $x$ along a latitude. Additionally, we assume the forcing profile to be a simple cosine in $x$, $h_{\rm eq}'=\Delta h_{\rm eq}\,f(t) \cos(kx)$, where $k$ is the horizontal wavenumber of the gravest planetary-scale wave and hence equal to $1/a$; see Figure \ref{fig:schem_hemi}. Here $f(t)$ is some function of time resembling the nature of instellation variation.

We seek solutions of the form
   $ h'(x,t) = \Delta h(t) \cos(kx) \label{eq:ansatz}$ corresponding to the gravest wave mode,
where $\Delta h(t)$ is the time-varying response to be determined. This solution exists under the assumption of pure global substellar to antistellar point flow, meaning that any other regional flow structures are neglected by this reduction. Despite this, we shall see later that numerical simulations match the analytical thermal response reasonably well. This is because the substellar to antistellar flow is divergent, and compared to the rotational component it is the main contributor to heat transport in the linear regime \citep{hammond2021rotational}. With the identity $gHk^2 = \tau_{\rm wave}^{-2}$, Equation \eqref{eq:lin_wave_pde} reduces to the following ODE for the planetary heat/height contrast,
\begin{align}
     \frac{\partial^2{\Delta h}}{\partial t^2} + \left( \frac{1}{\tau_{\text{rad}}} + \frac{1}{\tau_{\text{eff}}} \right) \frac{\partial \Delta h}{\partial t} + \left( \frac{1}{\tau_{\text{wave}}^2} + \frac{1}{\tau_{\text{rad}}\tau_{\text{eff}}} \right) \Delta h \nonumber \\
     = \Delta h_{\text{eq}} \left[\frac{1}{\tau_{\text{rad}}} \frac{\partial f(t)}{\partial t} + \frac{1}{\tau_{\text{rad}}\tau_{\text{eff}}} f(t) \right].
     \label{eq:so_ode}
\end{align}

We compare this equation with $\ddot{\Delta h} + \gamma \dot{\Delta h} + \Omega_{\rm o}^2 \Delta h =  F(t)$, which is the equation for a forced damped harmonic oscillator. The possibility of this reduction has been alluded to in \cite{Shamir2023} but not formally derived. This equation admits periodic solutions even without a sinusoidal forcing. The natural frequency and damping of the system are as follows,
\begin{align}
    \Omega_{\rm o} = \sqrt{\tau_{\rm wave}^{-2} + \tau_{\rm rad}^{-1}\tau_{\rm eff}^{-1}}, \nonumber \\
    \text{and } \gamma = \tau_{\rm rad}^{-1} + \tau_{\rm eff}^{-1}.
    \label{eq:nat_freq_damp}
\end{align} 

Once we have solved for $\Delta h$, we may substitute $h'$ back in Equation \eqref{eq:lin_mom_unsteady} to uniquely determine the magnitude of the velocity field, which would also show an oscillatory behaviour representing winds that grow and fade with changing intensity of instellation. We note that the velocity field varies as a sine function from the substellar point because of the gradient operator on $h'$. The highest and lowest temperature points (max/min $h'$), namely the hotspot and the coldspot, are therefore regions of stagnation, and the winds reach maximum speed near the day--night divide (i.e., the terminator). 


\subsection{Periodic forcing: The harmonic oscillator analogy}
\label{sec:variable_instellation}

\begin{figure*}
    \centering
    \includegraphics[width=0.85\linewidth]{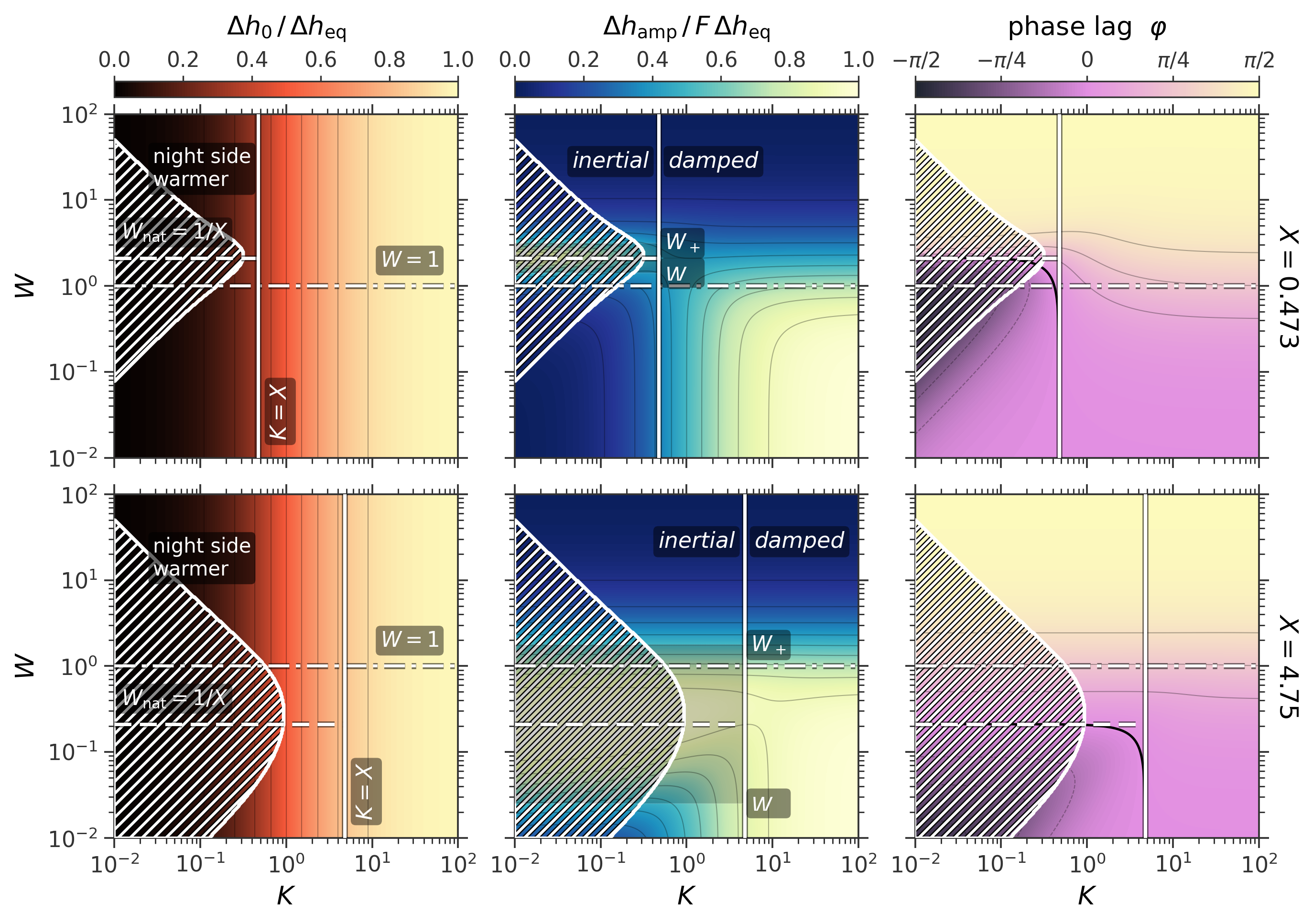}
    \caption{\textbf{A tidally locked planet behaves like a harmonic oscillator when subjected to time varying radiative forcing.} The three columns show the normalized time-mean height magnitude ($\Delta h_0/\Delta h_{\rm eq}$) or the constant irradiation response, the amplitude ($\Delta h_{\rm amp}/ F\Delta h_0$), and the corresponding phase lag of the response from the stellar forcing as in Equations \eqref{eq:so_mean}, \eqref{eq:so_amp} and \eqref{eq:so_phase}, respectively. The results are shown in the $K$--$W$ plane, where $K$ is the heat retention parameter, and $W = \omega \tau_{\rm rad}$ is the dimensionless forcing, for two different classes of planets, with the top row for weakly irradiated, smaller Earth-like planets and the bottom row for strongly irradiated gas giant-like planets. The vertical solid white line ($K = X$) represents separation between the wave dominated or inertial regime to the left of it, where the system response is second order and hence features resonances, and the drag/rotation dominated or damped regime to its right, where the system response reduces to a first-order, linearly damped relaxation towards the forcing. The dashed horizontal line corresponds to the dimensionless natural frequency of the system where the instellation variation ($\omega$) matches the wave frequency ($
    \sim 1/\tau_{\rm wave}$), causing resonant amplification of the contrast amplitude. This amplitude is at least half its maximum value anywhere in shading between the $W_+$ and $W_-$ lines, which have been constructed to identify potentially observable systems later in the paper. Above the dash-dotted horizontal line, the forcing frequency is too high for the system to respond significantly. The white hatched region represents system parameters for which the total contrast may be temporarily negative, implying a nightside temporarily warmer than the dayside.
}
    \label{fig:foso_theory}
\end{figure*}

We may understand the time-dependent response of the system as that of an oscillator in which gravity-wave adjustment provides the restoring force, while radiative relaxation and drag provide the damping. If the planetary atmosphere is suddenly illuminated with a flash of radiation, the dayside would heat up temporarily. This increases the height of the dayside which then adjusts itself by propagating throughout as gravity waves, which may interfere with one another sustaining global oscillations only if the effective drag (drag + rotation) is low (large $\tau_{\rm eff}$) and if the heat loss via radiative cooling is relatively slow (large $\tau_{\rm rad}$). The response is the same as the classical spring-mass-dashpot system subject to a delta kick.

In this section, we shall start focusing on systems with \textit{periodically} varying instellation, the central theme of this work. The response of such a system resembles a \textit{forced} damped harmonic oscillator. Even though the real nature of variability can be significantly complicated, we start with simpler forms. We assume the forcing to be a simple sine superimposed on a background value and set $f(t) = (1 + F\sin \omega t)$, where $\omega$ is the instellation variation frequency. There is a transient response which we may ignore here and look at the steady-state response. We obtain solutions of the form
$
   \Delta h(t) = \Delta h_{0} \;+\;  \Delta h_{\mathrm{amp}}\,\cos\bigl(\omega t - \varphi\bigr)
    \label{eq:case3_alt_ansatz}
$
to get,
\begin{equation}
    \frac{\Delta h_0}{\Delta h_{\mathrm{eq}}}
    = \left(1 + \dfrac{\tau_{\rm rad} \tau_{\rm eff}}{\tau_{\rm wave}^2}\right)^{-1} = \frac{1}{1+1/K},
    \label{eq:so_mean}
\end{equation}
\begin{align}
   A = \frac{1}{F} \cdot \frac{\Delta h_{\mathrm{amp}}}{\Delta h_{\mathrm{eq}}}  = \frac{\dfrac{1}{\tau_{\text{rad}}} \sqrt{\dfrac{1}{\tau_{\text{eff}}^2} + \omega^2}}{\sqrt{(\Omega_{\rm o}^2 - \omega^2)^2 + (\gamma \omega)^2}}  \nonumber \\ = \frac{\sqrt{K^2 + W^2 X^4}}{\sqrt{(1 + K - W^2 X^2)^2 + W^2 (X^2 + K)^2}},
   \label{eq:so_amp}
\end{align}
\begin{align}
    \text{and }\tan \varphi
    &= \frac{\omega\left[\gamma\,\tau_{\rm eff}^{-1} - \left(\Omega_{\rm o}^2 - \omega^2\right)\right]}
            {\left(\Omega_{\rm o}^2 - \omega^2\right)\tau_{\rm eff}^{-1} + \gamma\,\omega^2}
     = \frac{W\left(K^2 - X^2 + W^2 X^4\right)}{K\left(1 + K\right) + W^2 X^4}, \nonumber \\
    \text{equivalently}\quad
    \varphi &= \arctan\!\left[\frac{W\left(X^2 + K\right)}{1 + K - W^2 X^2}\right]
             - \arctan\!\left(\frac{W X^2}{K}\right).
    \label{eq:so_phase}
\end{align}

Here we have defined two new nondimensional parameters, the nondimensional instellation variation frequency $W=\omega \tau_{\rm rad}$ and the timescale ratio $X=\tau_{\rm wave}/\tau_{\rm rad}$. Thus, the dimensionality of the parameter space is reduced from seven dimensional parameters ($f, g, a, H, \tau_{\rm rad}, \tau_{\rm drag}$, and $\omega$) to just three nondimensional parameters $X, K$ and $W$. The nondimensional natural frequency of the system is $\Omega_{\rm o} \tau_{\rm rad} = \sqrt{(1 + K)/X^2}$. $X=\tau_{\rm wave}/\tau_{\rm rad}$ is the measure of how fast waves cross the planet relative to radiative relaxation. This parameter differentiates various planet types. For example, mildly irradiated Earth-like or other distant planets typically have $X<1$, while strongly irradiated hot Jupiter-like planets have $X>1$, depending on their relative wave speed ($=a/\sqrt{gH}$) which in turn is a function of the size, composition and atmospheric thickness of the planet. For this work we fix the values of $X$ at 0.473 and 4.745 to represent two classes of planets that lie on either side of the $X=1$ boundary; see Table \ref{tab:system_params}.


\subsection{Inertial and damped responses}\label{sec:in_dam_resp}

At this point, it is useful to determine one last scale, the characteristic timescale on which the height field itself evolves. When we compare the scales of the unsteady ($\partial h'/\partial t$) and the height advection (mass flux, $H\nabla\cdot\mathbf{u}'$) terms in Equation \eqref{eq:lin_cont}, we obtain the following timescale for the flow in general,
\begin{equation}
    \frac{\Delta h}{\tau_{\rm o}} \sim \frac{HU}{a} \implies \tau_{\rm o} \sim \tau_{\rm adv} \cdot \frac{\Delta h}{H} = \frac{\tau_{\rm wave}^2}{\tau_{\rm eff}} \label{eq:scale_time}
\end{equation}
where $\tau_{\rm adv} = a/U$ is the advection timescale, or the timescale over which winds traverse across the planet. The unsteady term ($\partial \mathbf{u}'/\partial t$) in the momentum Equation \eqref{eq:lin_mom_unsteady} is significant whenever its scale is much greater than that of the effective drag term, $U/\tau_{\rm o} \gg U/\tau_{\rm eff}$, or $\tau_{\rm o} \ll \tau_{\rm eff}$. When we substitute the timescale $\tau_{\rm o}$ from Equation \eqref{eq:scale_time}, the condition becomes
    $\tau_{\rm wave} \ll \tau_{\rm eff},$
which is the same condition as inequality \eqref{eq:condt_inertial}. This is expected because the unsteady and advective terms are components of the material derivative and hence have the same scaling. When the condition is satisfied, rotation and drag are weak and cannot damp gravity wave perturbations from reverberating across the planet, and the height field simply relaxes towards the forcing giving rise to linearized inertial physics. When waves dominate new behaviour emerges through constructive or destructive interference leading to resonances, as we shall see further. 

Figure \ref{fig:foso_theory} shows the time-mean height magnitude, its amplitude, and phase of planetary response for the two different values of $X$ (= 0.473, 4.75)) and will be compared with simulations in the upcoming section. So far as the constant part of the solution, Equation \eqref{eq:so_mean}, is concerned, it is exactly what has been derived from scaling laws in Section \ref{sec:steady_scaling} in Equation \eqref{eq:scale3}, which is expected in the absence of variable irradiation. The dynamical response comes from the time varying effect. The first thing to note is the region in $K$--$W$ space where the inertial physics dominates. This is demarcated by the solid white line $K=X$, which is equivalent to the condition in Equation \eqref{eq:condt_inertial} in dimensionless form. To the left of the white line, $K < X$ or $\tau_{\rm wave} < \tau_{\rm eff}$, i.e.\ gravity waves traverse the planet before they are damped. This supports a wave-driven resonance. Conversely, for $K > X$ (right of the white line) the wave timescale is too large and the drag too strong. The solid white line thus separates a wave-active inertial regime from a drag-dominated regime in which the restoring force is ineffective and the planet merely relaxes towards the instantaneous forcing. Outside of the inertial regime, the unsteady momentum term can be ignored, establishing a balance between pressure gradient and the effective drag. This reduces the system further to first order, where it merely represents a periodically forced and linearly damped system that does not exhibit any resonance phenomena. Since this happens to the right of the $K=X$ line, the first-order theory will be valid for the entire $K$--$W$ parameter space when $X \to 0$, pulling the white line all the way to the left. Now, Equations \eqref{eq:so_amp} and \eqref{eq:so_phase} reduce to $\Delta h_{\rm amp}/(F \Delta h_{\rm eq}) = \left[(1+1/K)^2 + W^2\right]^{-1/2}$, and $\tan\varphi = W/(1+1/K)$, which is the response of a system with no restoring force at all.


The second thing to note is that the inertial regime ($K \ll X$) features a local maximum in frequency space corresponding to the resonance. In this limit when $K \to 0$, we find the following relation for the dimensionless natural frequency,
\begin{equation}
    W_{\rm nat} = 1/X \text{ or } \omega_{\rm o} = 1/\tau_{\rm wave}.
    \label{eq:res_cond}
\end{equation}
This condition holds when the instellation variation time matches the wave timescale of the planet. Thus, given the system is purely inertial ($K \ll X \text{ or } \tau_{\rm wave} \ll \tau_{\rm eff}$), forcing the system at its wave frequency produces resonant amplification of the response. This can also be concluded from Equation \eqref{eq:nat_freq_damp} when $\tau_{\rm eff}, \tau_{\rm rad} \to \infty$. The correspondence can be seen at low values of $K$ for respective $X$, where the white dashed lines representing the natural frequency asymptote to become parallel to the x-axis close to $W=1/0.473\sim2$, and $1/4.745\sim0.2$ for the two values of $X$, respectively. 

It is useful to attach a width to the resonance, so that a system can be called near-resonant if it falls within it. In the same $K \to 0$ limit Equation \eqref{eq:so_amp} reduces to $A = WX^2/\sqrt{(1-W^2X^2)^2 + W^2X^4}$, which peaks at $A = 1$ exactly at $W = W_{\rm nat} = 1/X$, for every $X$. When we set $A = 1/2$, we obtain the positive roots of the corresponding quadratic, which bound the \emph{half-amplitude window},
\begin{equation}
    W_{\mp} = \frac{\sqrt{3X^2+4} \mp \sqrt{3}\,X}{2X}.
    \label{eq:half_amp_window}
\end{equation}
The width of the half-amplitude window is universal, $W_+ - W_- = \sqrt{3}$ for every $X$. Also, $W_-W_+ = 1/X^2$, making $W_{\rm nat}$ the geometric mean of the bounds, and hence the window is centred on $W_{\rm nat}$ in $\log W$ space. The window is drawn on the amplitude panels of Figure~\ref{fig:foso_theory} and is later used in Section~\ref{sec:kw_space} to decide which observed systems are near resonance.

The linear response to variable instellation is
$\Delta h(t) = \Delta h_0 + \Delta h_{\rm amp}\cos(\omega t - \varphi)$, with the
steady and oscillating parts given by
Equation~\eqref{eq:so_mean} and
Equation~\eqref{eq:so_amp}, respectively. The day--night contrast therefore passes through
zero and becomes negative at some phase of every cycle whenever the oscillating part
exceeds the steady one,
\begin{equation}
    F\,A > \frac{K}{1+K},
    \label{eq:inversion}
\end{equation}
so that the nightside is briefly warmer than the dayside. The hatched regions in Figure \ref{fig:foso_theory}
mark where Equation~\eqref{eq:inversion} holds at $F = 0.5$, the fractional semi-amplitude of forcing that is used in all simulations that follow in the next sections. Since $A \le 1$ everywhere, a
necessary condition is $K < F/(1-F) \sim 1$, so the reversal is confined to the weakly
damped corner overlapping the region of resonant amplification. We can think of the reversal as the planet being pumped close to the natural wave frequency of the planet, which produces a global standing wave that periodically shifts peaks and troughs between the substellar and antistellar points and makes the nightside occasionally warmer. The observational implications of this nightside warming are described later in the context of phase curves in Section~\ref{sec:phase_curves}.

Finally, $\varphi$ is the lag of the oscillating height contrast behind the instellation $f(t)$, not the hotspot offset from the substellar point, and it lies within $-\pi/2 < \varphi < \pi/2$. For slow forcing ($W \to 0$) the contrast follows the instellation and $\varphi \to 0$, whereas for fast forcing ($\omega \gg \Omega_{\rm o}$) the lag approaches $\pi/2$ because the oscillator cannot follow the forcing. In the inertial regime ($K < X$), $\varphi$ is negative below $W = \sqrt{X^2 - K^2}/X^2$ (black line in Figure \ref{fig:foso_theory} phase plots), which is close to $W_{\rm nat}$ when $K \ll X$, so the contrast peaks before the instellation does. This lead arises because the forcing in Equation~\eqref{eq:so_ode} contains $\dot{f}$, the rate of change of the forcing, which dominates the forcing $f$ when $1/\tau_{\rm eff} \to 0$. 


Thus, the response of tidally locked planets to time-variable irradiation is that of a damped harmonic oscillator, in which gravity-wave adjustment provides the restoring force while radiative relaxation and drag provide the damping. There are three conditions that control the response boundaries, namely the separation between inertial and damped response ($\tau_{\rm wave} \sim \tau_{\rm eff}$), the condition of resonant amplification ($\tau_{\rm wave} \sim T_{\rm instel}$), and whether the forcing period is slower than the radiative timescale or not ($\omega \sim 1/\tau_{\rm rad}$). In the following sections we test our theory through numerical simulations of systems that span the first two criteria, because the dynamics at these two criteria has not been studied for tidally locked planets. For these systems, we investigate the role of nonlinearities and how the system behaviour departs from the linearized theory presented above. We also discuss the consequence of the last condition briefly in Appendix \ref{app:long-taurad}.

\subsection{Assumptions and limitations of the analytical framework}\label{sec:assu_lim}

The analytical results rest on several simplifying assumptions. First, the effective drag approximation combines the Coriolis term and Rayleigh drag into a scalar $\tau_{\rm eff}^{-1} = 2\Omega\sin\phi_0 + \tau_{\rm drag}^{-1}$ (Equation~\ref{eq:tau_eff}), which adequately estimates global heat redistribution \citep{perez2013atmospheric}, but discards the directional information needed to predict the eastward versus westward direction of hotspot oscillations. The simulations retain the full Coriolis term and reproduce this transition. Second, we assume a cosine ansatz for the trial solution $h'(x,t)=\Delta h(t)\cos(kx)$ that is symmetric about the substellar point implying purely substellar to antistellar point flow, so the theory does not predict any spatial hotspot displacement. The actual offset arises from the 2D Matsuno--Gill response \citep{HammondPierrehumbert2018,PierrehumbertHammond2019}, which is not included in this 1D framework. Third, we retain only the gravest wave mode $k=1/a$ spanning the entire planet. At fast rotation, the Rossby deformation radius $L_d = c/f \ll a$ and higher harmonics become energetically important, causing the theory to overestimate the planetary response (see simulation results later). Fourth, the meridional gradients are simplified by assuming periodic boundary conditions along longitudes, reducing the problem to a 1D zonal diffusion equation. This equatorial-strip approximation breaks down at fast rotation where off-equatorial Rossby wave structure becomes significant. Finally, the term $\mathbf{R}$ (Equation~\ref{eq:SW_mom}), which is responsible for equatorial superrotation \citep{ShowmanPolvani2011}, is omitted from the linearized theory. The theory therefore cannot predict the equatorial jet, but the large-scale pressure-drag balance remains a reasonable approximation even in its presence.

\begin{figure*}
    \centering
    \includegraphics[width=0.8\linewidth]{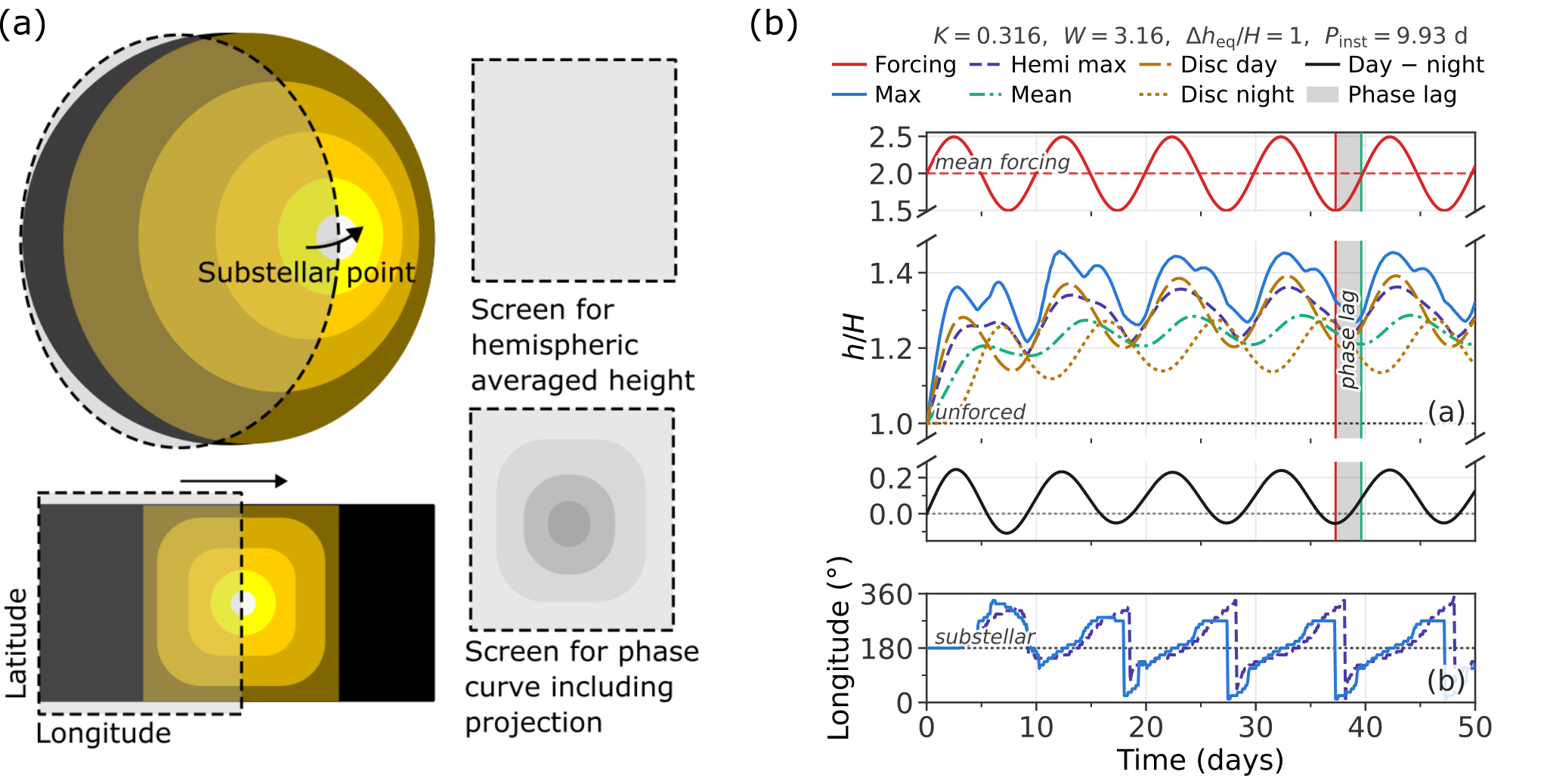}
    \caption{\textbf{Problem setup and diagnostics for 2D shallow water simulations.} \textbf{(a) Typical equilibrium height distribution ($h_{\rm eq}$) of a tidally locked planet in the absence of winds, towards which the planet is radiatively forced.} The same field is shown in two representations, spherical and rectangular. The substellar point lies at the centre of the dayside. The rectangular representation additionally shows the latitude, longitude and the unforced nightside in black. To the right are the hemispheric screens used to calculate diagnostics such as the hemispheric-averaged height and phase curves (see text for details). The screen for hemispheric-averaged height is plain, while the second screen represents the hemisphere visible along the line of sight of the observer and includes a cosine projection with a maximum at the substellar point to account for ray tracing effects needed for phase curve predictions (further details in Section~\ref{sec:phase_curves}). \textbf{(b) Nonlinear shallow-water response to time-varying instellation for a representative Earth-like simulation showing the principal diagnostics.} Sub-panel (a) shows the relevant normalized height fields, $h/H$, namely the forcing (red), its mean (red dashed), the global maximum of the response (blue), its hemispheric/disc-averaged maximum (purple dashed) and the global mean (green), the disc-averaged dayside (dashed golden), the disc-averaged nightside (dotted golden) and the net day--night contrast (black). The dotted black line is the unforced nightside height $h/H = 1$. Panel (b) shows the time evolution of hotspot locations represented by the longitude of the global and hemispheric maxima, with the substellar point as a dotted line. For this simulation $K = 0.316$ (rotation period $26.4$~d), $W = 3.16$ and the forcing strength $\Delta h_{\rm eq}/H = 1$ modulated at the instellation period $T_{\rm instel} = 9.93$~d. Only the first 50~d ($\sim$5 cycles) of the run are shown.}
    \label{fig:schem_hemi}
\end{figure*}

\begin{figure}
    \centering
    \includegraphics[width=0.9\linewidth]{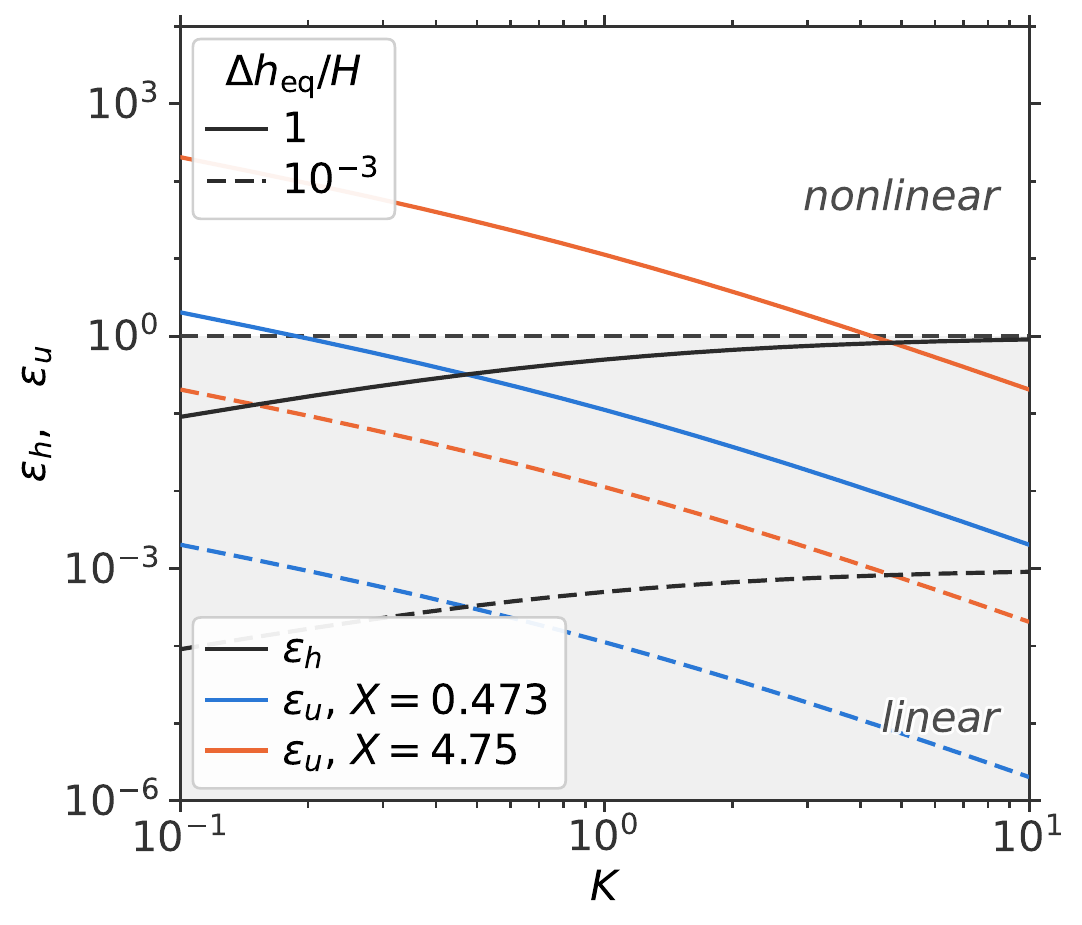}
    \caption{ \textbf{Smaller Earth-like planets primarily span the linear regime, while gas giant-like cases experience momentum nonlinearity for strong forcing.} Magnitude of height and momentum linearity constants $\varepsilon_h$ and $\varepsilon_u$ from inequalities \eqref{eq:lin_condt2} and \eqref{eq:lin_condt1}, respectively, as functions of $K$ for two different planet classes assumed in this work, $X=\tau_{\rm wave}/\tau_{\rm rad} = 0.473$ for smaller Earth-like and 4.745 for gas giant-like planets. Solid and dashed lines correspond to strong and weak forcing, respectively. The strongly forced gas giant set is markedly nonlinear in momentum because $\varepsilon_u \gg 1$ for low $K$ values, while the former lies reasonably within linear limits for both height and momentum with $\varepsilon_h,\varepsilon_u \lesssim 1$.}
    \label{fig:regime_constrast}
\end{figure}

Despite these limitations, the large-scale thermodynamic response, including the day--night contrast, oscillation amplitude, and phase lag, is well captured by the theory, as confirmed by the simulations in the following section.

\section{Numerical simulations}
\label{sec:simulations}

Nonlinear hydrodynamical simulations help us identify the limitations of theoretical results by relaxing the assumptions made to reduce the equations to simpler forms. By their inherent two-dimensional nature they also capture global flow patterns and corresponding variability which cannot be studied using the reduced-order theory above. In this section, we turn to full computational solutions to our shallow water equations and compare the results with our previous theoretical predictions.

\subsection{Framework}

Equations (\ref{eq:SW_mom}) and (\ref{eq:SW_mass}) are solved numerically using Dedalus3 \citep{burns2020dedalus}, a flexible pseudo-spectral code for symbolic equation entry. We discretize the 2D domain on a spherical basis and evolve the system using an SBDF2 timestepper. The simulation resolution is $32 \times 16$ (longitude $\times$ latitude), which, while low, sufficiently captures the core physical phenomena. This resolution corresponds to a minimum resolvable horizontal scale of $\sim$1000\,km for an Earth-like planet. This scale is adequate for the dominant wavenumber-1 day--night response but insufficient to resolve structures smaller than the equatorial Rossby deformation radius at the fastest rotation rates explored. We have conducted grid convergence studies for constant instellation cases and they show that the lowest resolution used here is sufficient to capture the linearized dynamics which is the central theme of the paper. We have previously verified our code by reproducing results from existing literature \citep{perez2013atmospheric, penn2017thermal, ohno2019atmospheres} for the work on variably irradiated asynchronous planets presented in \cite{Banik2025}, which is what has been extended here for tidally locked ones.

The tidally locked height field is relaxed towards an equilibrium profile $h_{\text{eq}}$ as shown in Figure~\ref{fig:schem_hemi}(a), the intensity of which is periodically varied. The form of the equilibrium height field is given by

\begin{align}
    h_{\rm{eq}}(\lambda,\phi,t) = H + \Delta h_{\rm eq}\cdot\frac{\mathcal{R} + |\mathcal{R}|}{2} \left(1 + F \sin \omega t \right) \nonumber
    \\
    \text{where} \hspace{2mm} \mathcal{R} = \cos \lambda \cos\phi.
\end{align}

Here, $H$ is the nightside thickness, $\Delta h_{\rm eq}$ is the forcing day--night contrast, and $F$ and $\omega$ are the instellation fluctuation amplitude and frequency. The factor $(\mathcal{R} + |\mathcal{R}|)/2$ isolates hemispheric irradiation \citep{dobrovolskis2009insolation}.


\subsection{Diagnostics}
Each run produces timeseries of global height and velocity fields. These are post processed to obtain the quantities that can be compared with the theory above. From the global height field we obtain the timeseries of a range of quantities pertaining to the height field, namely the \textit{maximum} height, the global \textit{mean} height, the disc-averaged dayside height, similarly for the nightside, and the difference or the contrast. We also compute the disc-averaged height centred at each longitude to get an azimuthal variation of the height field, which we call the \textit{hemispheric average} height; see Figure \ref{fig:schem_hemi}(b). The time-mean magnitude of planetary response is calculated by taking disc-averaged contrast timeseries after the simulation is out of its transient evolution phase and relaxed to an oscillating steady state, typically over the last three oscillation cycles of every simulation. The \textit{amplitude} is half the modulus of the difference between the peak and trough of the same. The statistical steady state is identified by tracking the time-averaged kinetic energy of the flow to have reached a constant value. Finally, the \textit{phase} is obtained by calculating the temporal offset between the troughs in instellation and \textit{mean} height and converting it to radians; see third panel in Figure \ref{fig:schem_hemi}(b). The reason for choosing the mean height here is that it has a smoother sinusoidal structure than maximum height, especially for inertia dominated cases where interacting waves interfere to create complicated responses. 

We also obtain another estimate of the global height field, the \textit{hemispheric averaged} height. It is calculated as an average of the height field over a longitudinal extent of $\pi$ radians, the face of the planet centred at each individual longitude, and is a more accurate metric when comparing with observations that average the light coming from the exposed planetary hemisphere. The projection used for this calculation is shown in Figure \ref{fig:schem_hemi}(a). We also obtain the longitudinal location of maximum height and the maximum hemispheric averaged height to track the motion of the real and observationally apparent hotspots. 

\subsection{Exploration parameters}
\label{sec:exploration}

\begin{table*}
\centering
\caption{Physical parameters and timescales corresponding to the simulations presented. The table shows a total of 100 simulations, each of which are conducted at two different values of $\Delta h_{\rm eq} = 10^{-3} H$ and $H$, representing the weak and strong forcing regimes of \cite{perez2013atmospheric}, leading to linear and nonlinear behaviour described above, resulting in a total of 200 simulations.}
\label{tab:system_params}
\begin{tabular*}{\textwidth}{@{\extracolsep{\fill}}llll}
\toprule
\textbf{Parameter} & \textbf{Symbol} & \textbf{Earth-like} & \textbf{Gas giant-like} \\ 
\midrule
Surface gravity & $g$ & $9.81\,\text{m}\cdot\text{s}^{-2}$ & $10\,\text{m}\cdot\text{s}^{-2}$ \\
Scale height & $H$ & $100\,\text{m}$ & $4 \times 10^{5}\,\text{m}$ \\
Radiative timescale & $\tau_{\text{rad}}$ & $5\,\text{days}$ & $0.1\,\text{days}$ \\
Instellation variation & $T_{\rm instel}$ & $3.14\text{--}314\,\text{days } (\times 5)$ & $0.06\text{--}6.28\,\text{days }  (\times 5)$ \\
Wave timescale (derived) & $\tau_{\text{wave}}$ & $2.365\,\text{days}$ & $0.4745\,\text{days}$ \\
Wave-to-radiative timescale ratio (derived) & $X$ & $0.473$ & $4.745$ \\
\addlinespace
\midrule
\textbf{Scenario} & \textbf{Rotation period ($P_{\text{rot}}$)} & \textbf{Drag timescale ($\tau_{\text{drag}}$)} & \textbf{Number of runs} \\ 
\midrule
\textbf{A: Earth-like} & & & \\
- Batch 1 & $140\,\text{days}$ & $0.11\text{--}22.373\,\text{days } (\times 5) $ & 25 \\
- Batch 2 & $0.7\text{--}140\,\text{days}$ & $22.373\,\text{days } (\times 5)$ & 25 \\
\addlinespace
\textbf{B: Gas giant} & & & \\
- Batch 3 & $282\,\text{days}$ & $0.22\text{--}45\,\text{days } (\times 5)$ & 25 \\
- Batch 4 & $1.42\text{--}282\,\text{days}$ & $45\,\text{days } (\times 5)$ & 25 \\
\bottomrule
\end{tabular*}
\end{table*}

We use two different magnitudes for the base relative forcing $\Delta h_{\rm eq}/H = 10^{-3}$ and $1$, respectively to span linear and nonlinear behaviour. Figure \ref{fig:regime_constrast} plots the linearity constants $\varepsilon_h$ and $\varepsilon_u$ as a function of $K$ for two different forcing magnitudes $\Delta h_{\rm eq}/H$, and two different classes of planetary parameters, the first corresponding to smaller Earth-like planets for which $X=0.473$ \citep{penn2017thermal,ohno2019atmospheres,Banik2025}, and the other corresponding to typical close-in gas giants for which $X=4.745$ \citep{perez2013atmospheric}. We see that the momentum condition is decisively nonlinear for the latter group at strong forcing for most values of $K$ (orange solid line), while the other cases are relatively linear. The differences in their response to variable irradiation show up distinctly when hotspot locations are analysed later in the paper. For each planet class, we simulate 25 planets that span the entire nondimensional $K$--$W$ parameter space sampling 5 equally log-spaced values between $10^{-1}$ and $10^1$;  see individual panels in Figure \ref{fig:earth_analog}.

The fractional semi-amplitude of the forcing $F$, which signifies the amount by which the stellar irradiation varies, is kept at 0.5 to elicit significant response given its linear scaling with the response. We run these in two batches with $X=0.473$ and 4.745, defining different ratios of radiative and wave adjustment strengths. The first corresponds to smaller Earth-like planets with slow wave propagation, while the latter uses the parameter set in \cite{perez2013atmospheric} applicable to gas giants which are bigger and hence have faster wave propagation; see Table \ref{tab:system_params}. 

Here, we recall that in the linearized theory we have assumed the Coriolis and drag terms to be combined into the effective drag term for simplification. However, their effects on the global flow patterns are distinct because of the nature of the individual terms. The Coriolis term leads to latitudinal variations being stronger closer to the poles and weaker at the equator, while the drag depends only on the velocity and the constant $\tau_{\rm drag}$. For this reason, we vary $K$ by varying both rotation and drag individually, to study if contrast properties are actually significantly affected by them. Finally, we vary $W$ by changing the instellation variation frequency. All permutations included, we have a suite of 200 simulations ($25 \times 2 \times 2 \times 2$), the results of which are presented below. The details of the planetary parameters and total number of runs are given in Table \ref{tab:system_params}.

\subsection{Results}\label{sec:sim_results}

In the following sections, we present the simulated responses for various systems representing Earth-like ($\tau_{\rm wave} < \tau_{\rm rad}$) and gas giant-like ($\tau_{\rm wave} > \tau_{\rm rad}$) regimes. Figures \ref{fig:earth_analog} and \ref{fig:hot_Jupiter} show the time-mean, amplitude, and phase and also the degree of hotspot oscillation as seen in simulations. For the first two quantities, the mean ($\Delta h_{\rm disc}$) and amplitude ($\Delta h_{\rm disc,\,amp}$) of the time-varying contrast (black line) in Figure \ref{fig:schem_hemi}(b) are used. We additionally divide the amplitude by the fractional semi-amplitude of forcing $F=0.5$ to normalize it to any $F$ value. We also show regions of the parameter space where the nightside gets warmer than the dayside using white hatches. These are compared with theoretical results presented in Figure \ref{fig:foso_theory}.


\subsubsection{Typical Earth-like planets}
\label{sec:earth_like}

\begin{figure*}
    \centering
    \includegraphics[width=\linewidth]{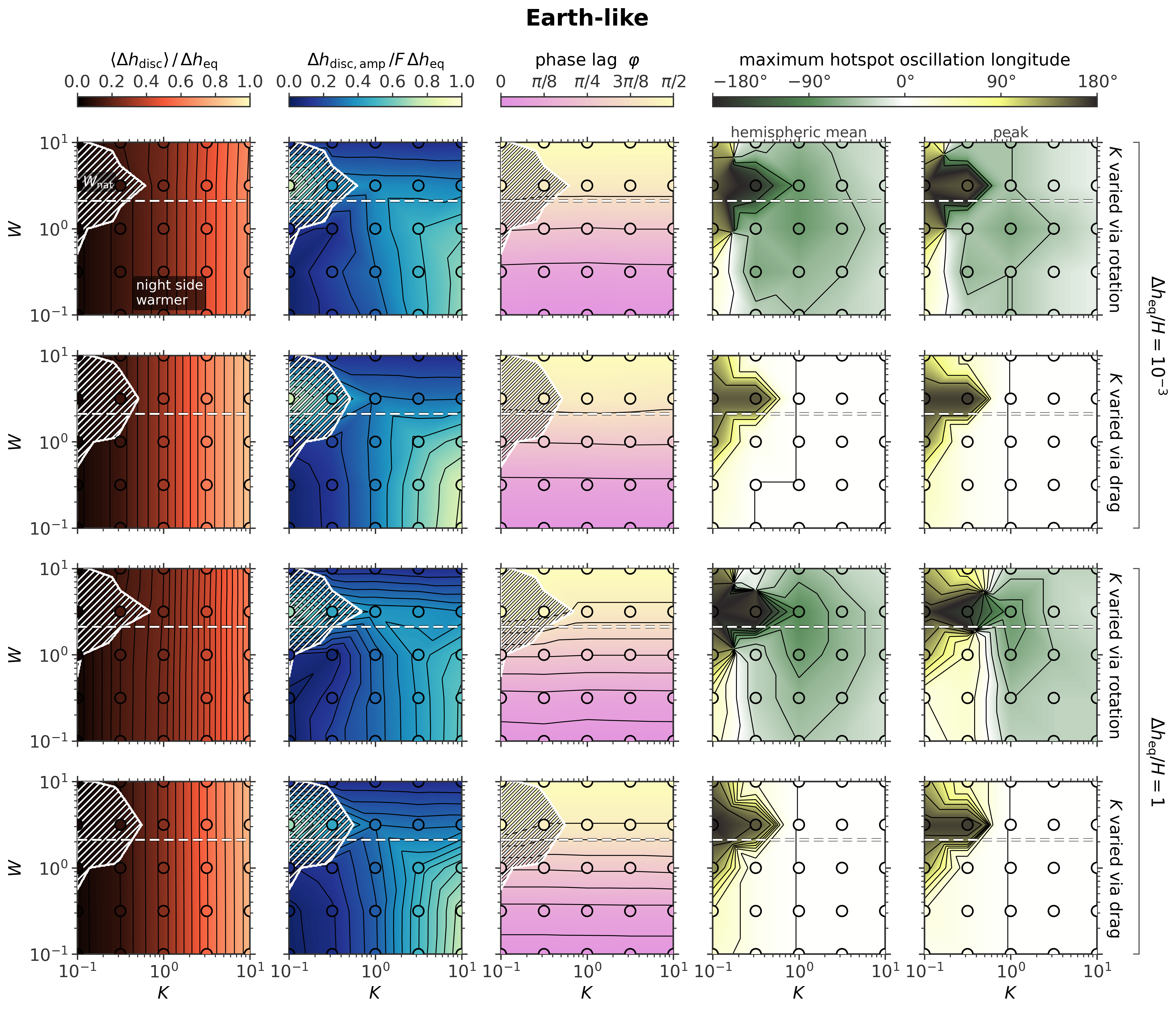}
    \caption{\textbf{Linear limit Earth-like simulations match theoretical predictions of planetary response to variable irradiation including resonant amplification.} Suite of simulation results for a tidally locked Earth-analogue planet, showing the normalized time-mean height magnitude ($\Delta h_{\rm disc}/\Delta h_{\rm eq}$), its amplitude ($\Delta h_{\rm disc,\,amp}/F\Delta h_{\rm eq}$), the phase lag from forcing ($\varphi$) and the maximum oscillating hotspot shift from the substellar point (both maximum height and hemispheric-averaged) in five columns, respectively. The colour schemes are identical to the ones used in theory plots in Figure \ref{fig:foso_theory}. The white dashed line represents the natural frequency of the planet in the inertial regime, which is also the frequency of propagating gravity waves $\sim 1/\tau_{\rm wave}$. The rows correspond to weak and strong forcing, with $K$ varied by rotation in one case and by drag in the other, each time holding the remaining parameter constant. Rotation and drag have a degenerate effect on the thermal response (first three columns appear not very different for individual cases) as predicted from scaling analyses, except for the maximum hotspot oscillation longitude (4th and 5th columns), which is a 2D behaviour only captured in simulations. The hotspot oscillations are also seen to maximize near the resonance which is also where the nightside is temporarily warmer in the forcing cycle as shown by white hatches. Additionally, hotspots show both eastward and westward oscillations with changing rotation, which is likely an impact of changing flow structures.}
    \label{fig:earth_analog}
\end{figure*}

In this section we present results for the Earth-analogue parameter set (Scenario A, Table~\ref{tab:system_params}; $g=9.81\,\mathrm{m\,s^{-2}}$, $a=6400\,\mathrm{km}$, $H=100\,\mathrm{m}$, $\tau_{\rm rad}=5\,\mathrm{days}$, $\tau_{\rm wave}=2.365\,\mathrm{days}$, $X=0.473$), which spans the linear regime for most of the $K$--$W$ space explored, with the exception of the lowest $K$ cases at $\Delta h_{\rm eq} = H$; see Figure~\ref{fig:regime_constrast}. 

\begin{figure*}
    \centering
    \includegraphics[width=\linewidth]{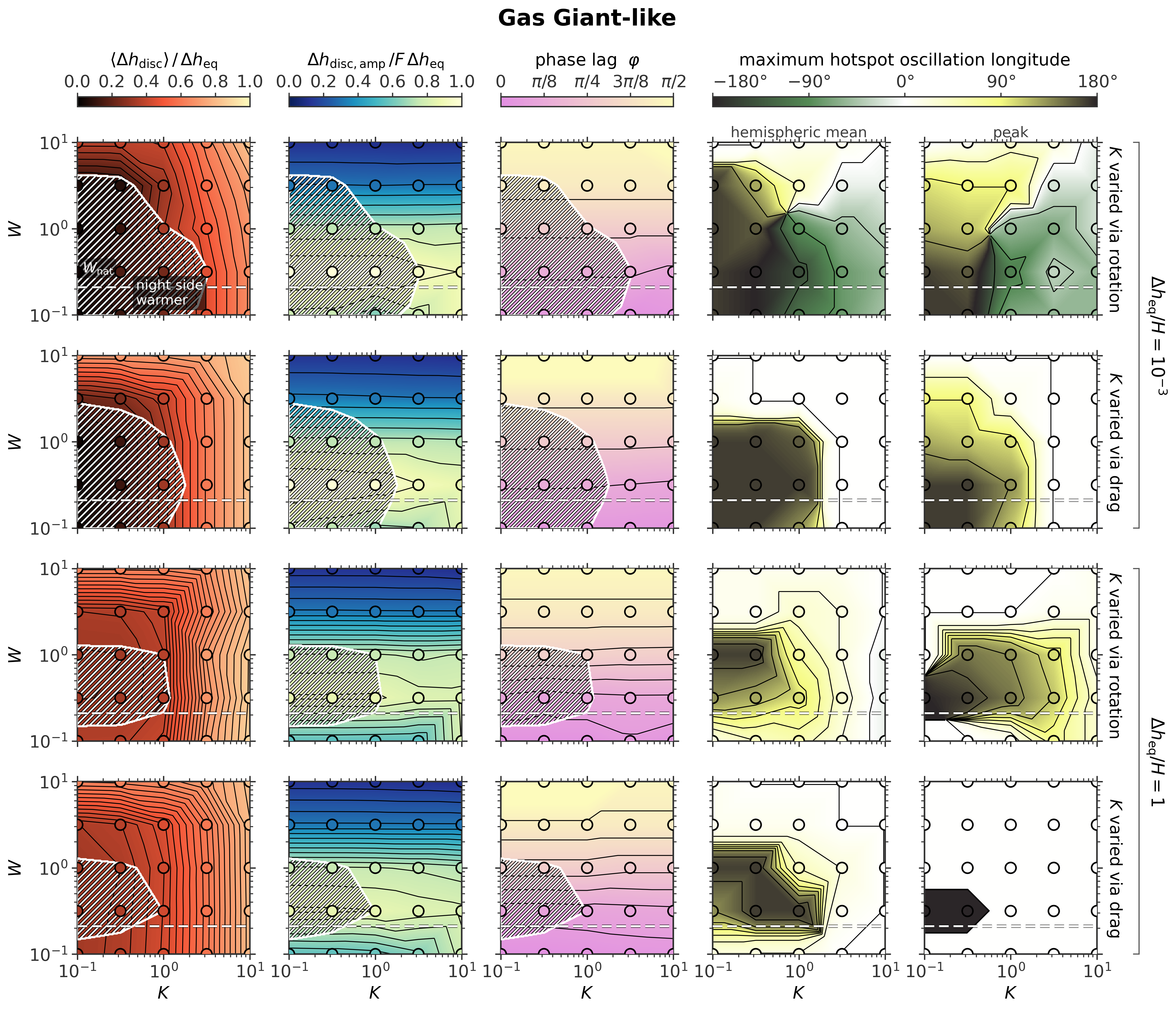}
    \caption{ \textbf{Strongly forced gas giant-like nonlinear limit simulations depart quantitatively from theoretical predictions, even though the resonant amplification and overall qualitative behaviour is preserved.} Suite of simulation results for a tidally locked gas giant planet, with the same quantities presented as in Figure \ref{fig:earth_analog}. Here, the differences between weak ($\Delta h_{\rm eq} = 10^{-3} \times H$) and strong ($\Delta h_{\rm eq} = 1 \times H$) forcing are stark. In particular, the amplitude and phase at strong forcing become independent of $K$, contrary to theory. The white hatch shows the region of relative nightside warming as obtained for individual runs, which also reduces in size for the nonlinear cases. Also, notably different is the loss of westward hotspot oscillations in the strongly nonlinear regime (compare column 4, rows 1 and 3). These departures are likely due to effects like wave-mean flow interactions which occur from nonlinearities in the system.}
    \label{fig:hot_Jupiter}
\end{figure*}

Figure \ref{fig:earth_analog} shows the simulated planetary responses for the above set. For the time-mean and amplitude parts of the response, we see a strong qualitative and a decent quantitative match with the theoretical results in Figure \ref{fig:foso_theory}, confirming that the assumption of pure substellar to antistellar point flow is sufficient to produce the correct trends for planetary response. This is because this flow structure represents the divergent part of the circulation which is known to be responsible for most of the heat transport in 3D atmospheres \citep{hammond2021rotational}. The region of resonance occurring at low $K$ shows a local maximum at $W=2\sim1/0.473$, close to the white dashed line corresponding to the theoretical prediction of the resonance. The region of the $K$--$W$ plane where the nightside gets occasionally warmer than the dayside, shown in white hatches, shows up around the region of resonance as predicted from theory. Here, at low $K$, the stationary contrast (first column) is small compared to the amplitude (second column) which experiences resonant amplification, taking the net contrast below zero and resulting in nightside warming.

The last column of Figure \ref{fig:earth_analog} plots the maximum amplitude of the hotspot oscillation, which is seen to correspond to the region of the $K$--$W$ parameter space where the theory predicts the resonant maximization of temporal variability. This is expected from the damped oscillator model of the planetary response, even though this phenomenon is a spatially varying one. This phenomenon is closely linked to the nightside warming, as we find the hotspot switching from being located at the substellar point to the antistellar point when the forcing frequency matches the wave timescale of the planet. Its direction, which distinguishes the drag-varying from the rotation-varying batches, is taken up in Section~\ref{sec:discussion}.

\subsubsection{Typical gas giants}
\label{sec:hot_jupiter}

In this section we present results for the gas giant parameter set (Scenario B, Table~\ref{tab:system_params}; $g=10\,\mathrm{m\,s^{-2}}$, $a=82{,}000\,\mathrm{km}$, $H=4\times10^{5}\,\mathrm{m}$, $\tau_{\rm rad}=0.1\,\mathrm{days}$, $\tau_{\rm wave}=0.4745\,\mathrm{days}$, $X=4.745$), taken from \cite{perez2013atmospheric}. Relative to the Earth-analogue, $X=4.745$ implies a significantly higher $\tau_{\rm wave}/\tau_{\rm rad}$ ratio. Only the momentum linearity condition is violated at $\Delta h_{\rm eq} = H$. The continuity equation remains linear throughout (see the red line in Figure~\ref{fig:regime_constrast}). 

Figure \ref{fig:hot_Jupiter} shows the planetary response to variable irradiation for the set of gas giant planets described above. When we compare with the theoretical plots corresponding to $X=4.745$, we find a significant qualitative match, although lower than the Earth-like cases, especially for strong forcing (bottom two rows). The resonant frequency has now moved down to a smaller value of approximately $W=0.2\sim 1/4.745$ obeying the $1/X$ rule from theoretical predictions in Section~\ref{sec:in_dam_resp}. The top two rows, for which $\Delta h_{\rm eq}$ is $10^{-3} \times H$, are in the linear regime, while the bottom two are not. As expected, the linear simulations match the theory better. For the bottom rows, the $\Delta h_0 / \Delta h_{\rm eq}$ plots have more averaged out values over the $K$--$W$ parameter space, although the contour lines show that the fundamental structure is preserved. Even the region of temporarily warmer nightside planets shown by the white hatch shrinks to a smaller area in the nonlinear regime. This is because several of our assumptions break down in nonlinear spherical simulations as mentioned in Section~\ref{sec:assu_lim}, and additional physics like wave-mean flow interactions come in. This is also seen in the amplitude plots with values averaged out for low $W$ values.   

We also observe that the phase offset has a decent match except where the response is expected to lead the forcing theoretically in Figure \ref{fig:foso_theory}. This is because the phase offset is calculated from the mean height rather than the day-night contrast and the algorithm considers all phases responses to lag the forcing, which is why negative values are not seen. We acknowledge that this could have been better calculated to show exact quantitative matches. However, for most of the simulations, the phase actually lags and hence matches the theoretical results fairly well. 


Overall, our simulations across Earth-analogue and gas giant parameter regimes show good agreement with theoretical predictions, which indicates that the drag and rotation contribute similarly to the overall thermal response and hence combining them into the effective drag term is not a poor assumption at first order. Linear regime simulations match theory more closely than nonlinear regimes, as expected. Despite their degenerate effect on the leading-order response, we also find that changing rotation and drag affect the time varying location of the planet's hotspot in different ways. The hotspot location affects the observational signatures of such planets, and we examine the hotspot behaviour in the next section.

\section{Global atmospheric variability}
\label{sec:discussion}

In this section we delineate the effect of varying irradiation on the atmospheric variability of a planet, specifically focusing on westward and eastward hotspot oscillations and speculate on the underlying mechanisms that cause these oscillations in linear and nonlinear regimes. Time-varying hotspot positions have been reported before, from two quite
different causes. In magnetohydrodynamic models a strong toroidal field
periodically reverses the equatorial jet, so the hotspot swings east and west on
the magnetic cycle rather than on any external clock
\citep{RogersShowman2014,RogersKomacek2014,Hindle2019,Hindle2021a,Hindle2021b}.
Purely hydrodynamic models produce displacements too, but they are intrinsic and
aperiodic, arising from the flow's own instability
\citep{Cho2003,KomacekShowman2020}; the observed westward offsets and
epoch-to-epoch changes are usually read in one of these two ways
\citep{Armstrong2016,Dang2018}. The oscillations below are of a third kind. They
are hydrodynamic yet strictly periodic, because the atmosphere is pumped
externally by the instellation, and their direction, as we shall see, is set by rotation rate and
nonlinearity rather than by field strength or by chance.

\subsection{Hotspot oscillations}
\label{sec:hovmoller}

From our simulations we find that the hotspot oscillates at the forcing frequency, moving away from the substellar point towards the antistellar point and back, so that varying irradiation produces thermal variability in both time and space. Its direction and longitudinal extent depend on both the rotation rate and the drag timescale, even though the theoretically derived thermal response parameters such as mean, amplitude and phase-lag are agnostic to these variables. For slowly rotating Earth-like and gas giant-like drag-varying batches, hotspot oscillations are purely eastward across all drag values. The longitudinal extent of oscillation varies over the $K$--$W$ parameter space with the maximum longitudinal shift ($180^\circ$) occurring at the location of the resonance and decaying in all other directions. Higher drag values lead to smaller oscillations, albeit justified by the dissipative nature of drag on wave or advection driven dynamics. The positive correlation between the maximum hotspot oscillation and the maximum planetary response amplitude at the resonance in the $K$--$W$ parameter space shows that the phenomena are connected, with the former likely also causing the latter. We expect this because both the resonance and hotspot oscillations result from inertial physics of the system. 

\begin{figure*}
    \centering
    \includegraphics[width=0.95\linewidth]{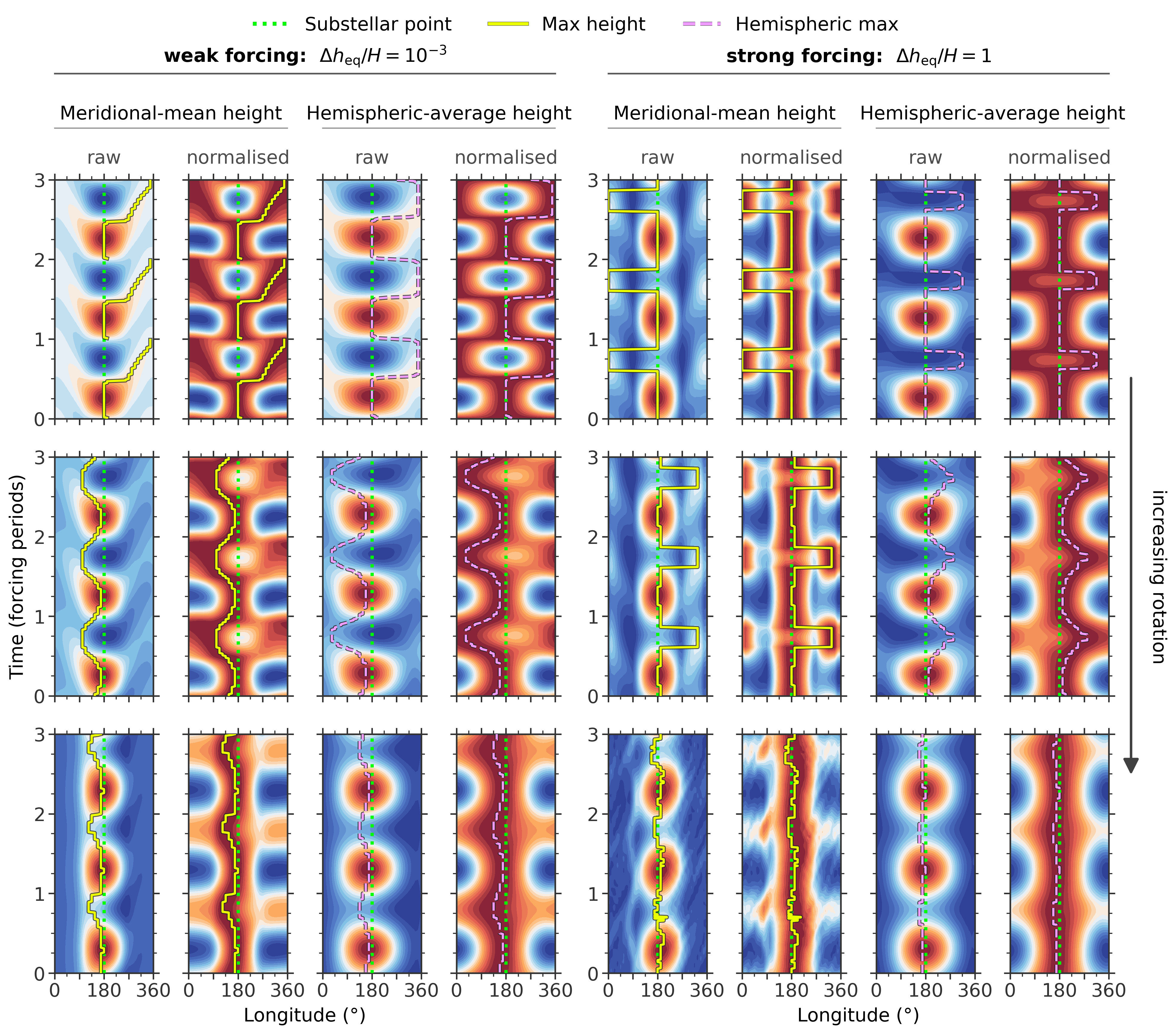}
    \caption{\textbf{Rotation and nonlinearities control the direction of hotspot oscillation on variably irradiated tidally locked planets.} Hovmöller diagrams or timeseries of hotspot location overplotted on background zonal height profiles for six gas giant planets with increasing rotation (282.9, 14.9 and 1.42 days from top to bottom, spanning weak to strong heat retention) in weakly forced linear ($\Delta h_{\rm eq} = 10^{-3}H$, left column) and strongly forced nonlinear ($\Delta h_{\rm eq} = H$, right column) regimes. Each planet has a collection of four individual panels that show different diagnostic quantities, namely the meridional mean height, the same normalized by the max height at each timestep to expose the instantaneous hotspot location (yellow lines), the hemispheric average height, and similarly normalized hemispheric-average with its own hotspot location (magenta lines). The dotted green line marks the substellar longitude. The colour bars for background height fields are dropped for better visualization. The time window shown spans the last 3 forcing periods of the respective simulations. Rotation at weak forcing experiences a switch from eastward to westward oscillations, while the oscillations remain purely eastward at strong forcing.}
    \label{fig:hovmoller}
\end{figure*}

On varying rotation, we find hotspot oscillations to be eastward at slow rotation and westward at fast rotation. The maximum amplitude of oscillation still occurs at the resonance for Earth-like cases. We note that the Earth-like system is primarily linear irrespective of forcing strength. In most cases, the hotspot oscillates on only one side of the substellar point. However, in some cases, especially close to the transition rotation rate, the hotspot oscillates across the substellar point both westward and eastward. In Figures \ref{fig:earth_analog} and \ref{fig:hot_Jupiter}, such cases are still represented by the maximum hotspot oscillation thereby omitting this information, but the actual complex hotspot oscillation is better visualized in Hovmöller diagrams \citep{hovmoller1949trough} shown hereafter in Figure~\ref{fig:hovmoller}. Finally, for the gas giant cases in the strongly forced nonlinear regime ($\Delta h_{\rm eq} = H$), we find that the oscillations are eastward even at fast rotation rates, so the behaviour is consistent across the entire parameter space. 

To visualize how rotation and nonlinearities alter the hotspot dynamics and to track the hotspot movement, we construct Hovmöller diagrams, which are longitude--time maps of the height field, for the three representative linear and nonlinear gas giant cases with changing rotation. Figure~\ref{fig:hovmoller} compares linear ($\Delta h_{\rm eq} = 10^{-3}H$) and nonlinear ($\Delta h_{\rm eq} = H$) runs at three rotation periods (282.9, 14.9, and 1.42 days), showing the last three forcing cycles of each simulation. Every panel shows the meridional mean height, averaged over a window $90^\circ$ wide about the equator, together with the hemispheric average height and the normalized version of each. There are two hotspot diagnostics overlaid on every panel, the location of maximum height or the hotspot (yellow line) and the hemispheric-maximum hotspot (magenta line). The dotted green line marks the substellar longitude. We note that the hemispheric average hotspot does not follow the same path as the meridional mean hotspot, because the two hotspots come from different forms of averaging applied to the same height field. The latter is a better indicator of the observable hotspot position owing to its projected face average (see Figure \ref{fig:schem_hemi}(a)). 

The first thing to note is that the planetary response is strongly modulated by the instellation in all cases. This is because the gas giant parameter set has a short radiative timescale of 0.1 days. Second, the normalization carried out for each individual timestep picks out the instantaneous height maxima, which makes the hotspot location visible. The absolute hotspot location matches the maxima of the normalized meridional mean height, while the hemispheric-maximum hotspot matches the normalized hemispheric average height. The Hovmöller diagrams add the shape of the hotspot excursion to the extent already shown in Figures~\ref{fig:earth_analog} and \ref{fig:hot_Jupiter}, in particular for the cases near the transition rotation rate, where the hotspot crosses the substellar meridian in both directions within a single forcing cycle.

\subsection{Role of rotation} \label{sec:role_of_rotation}

From the simulation results above, we note that the reflection of the resonance on the height response amplitude is rather minimal, but the hotspot oscillation is a more sensitive indicator. The resonance, if significant, always occurs at small values of rotation and drag (left-hand side of the $K$--$W$ plane). From linearized theory we know that strong rotation inhibits the transport of heat from the day to the nightside via waves (see also simulations of \citealp{Tan2020}), and so does strong drag. It is thus expected that the resonance and consequently hotspot oscillations will be the strongest when the wave propagation is least inhibited.

To understand the role of rotation let us consider slow rotation with \textit{constant} instellation first. In this case, the day--night radiative forcing creates a standing equatorial Kelvin wave pattern, akin to the gravest gravity wave mode, that closely resembles the equilibrium height field \citep{Gill1980,ShowmanPolvani2011}. Since the Kelvin wave has an eastward propagating phase velocity, on varying the instellation the system responds to the forcing by letting the Kelvin mode travel eastward until the instellation comes back to a maxima. Thus the point of maximum height, the hotspot, travels eastward and gets reset to the substellar point towards the end of the forcing cycle; see Figure \ref{fig:hovmoller}, yellow line in the top left panels. The slowest rotation case is extremely close to the non-rotating case, where gravity wave propagation is isotropic from the substellar to the antistellar point, which is what contributes to nightside warming as obtained theoretically in Section~\ref{sec:in_dam_resp}. This is why we see the height fields to be almost symmetric about the substellar point even though the hotspot oscillation points to the east as a result of the eastward propagating weak Kelvin wave mode being excited by the instellation forcing.

As the rotation rate is increased (second and third rows of the left-hand side panels of Figure \ref{fig:hovmoller} representing the weakly forced linear regime), the Coriolis force becomes stronger. This causes another off-equatorial wave mode, the Rossby wave, to be excited. The height fields are more asymmetric with the height maxima to the west of the substellar point (see second panel). The Rossby wave mode is westward propagating \citep{Matsuno1966,Gill1980} and may justify the westward oscillation observed at fast rotation rates.

This two-wave account, however, cannot be the whole story. In principle both the Kelvin and Rossby waves are excited at all rotation rates according to linearized $\beta$-plane theory \citep{Matsuno1966,Gill1980}. A change in $\Omega$ on the $\beta$-plane only rescales the two-wave solution in space and time and does not change the relative amplitude of Kelvin versus Rossby wave content, because that ratio is fixed by the forcing shape and damping alone in the scaled coordinates. Moreover, our system is on a sphere, where \cite{Shamir2023} show that further modes, namely the eastward and westward inertio-gravity waves and the mixed Rossby--gravity wave, are also present, each with its own direction of propagation, excited to different degrees by the nature of the forcing imposed, and differently sensitive to rotation and drag even where the height contrast is not. The exact mechanism for the switch in direction of hotspot oscillation is thus not clear, and we offer the Kelvin--Rossby argument above as a plausible reading rather than a demonstration. The mechanism will be investigated in a future study.

\subsection{Role of nonlinearities}
\label{sec:role_of_nonlinearities}

We now study the role of nonlinearities, the conditions for which are given in Equations~\eqref{eq:lin_condt2}~and~\eqref{eq:lin_condt1}. Simulations representing the nonlinear gas giant regime ($\Delta h_{\rm eq} = H$) are based on these conditions. However, the conditions themselves are agnostic to the forcing frequency. We have noted that a high forcing frequency can play a role in the local rise of nonlinearities, but we do not resolve small scales here and hence the nonlinear behaviour is restricted to its effect on the global scale features. We emphasize that nonlinear behaviour is not synonymous with wave-driven behaviour, even though both formally arise from the same inertial terms in the momentum equations.

In the nonlinear regime at low rotation (first row, right-hand side panels of Figure~\ref{fig:hovmoller}), we see the onset of a switching behaviour of the hotspot between the substellar and antistellar points, in contrast to the gradual eastward drift seen in the corresponding linear simulations. This hints at the presence of higher-order interactions that are absent in the linear regime \citep{ShowmanPolvani2010,ShowmanPolvani2011}. The exact mechanism responsible is not yet clear and warrants further investigation. However, the hotspot still oscillates eastward in this regime, similar to the linear case.

At intermediate rotation (second row, right-hand side panels), the hotspot oscillations continue to occur in the eastward direction, in contrast to the westward oscillation seen in the corresponding linear case. We suggest that the suppression of westward oscillations in the nonlinear regime is a consequence of the wave-mean-flow interaction mechanism of \citet{ShowmanPolvani2010,ShowmanPolvani2011}, \citet{tsai2014three} and \citet{HammondPierrehumbert2018}. In that mechanism, the equatorward eddy momentum flux convergence associated with the standing Kelvin--Rossby wave pattern spins up a persistent eastward equatorial superrotating jet. Once established, this jet Doppler-shifts the stationary forced wave pattern to produce an eastward hotspot offset in the absence of instellation variation \citep{HammondPierrehumbert2018}. In the linear regime, by contrast, the wave-driven momentum flux convergence is weaker and is unable to spin up or maintain an equatorial jet against drag. Without a jet to Doppler-shift the wave pattern, the Rossby-eddy pattern is free to dominate at fast rotation, producing the westward oscillations described above. We note, following \citet{Nicolas2026}, that the strength of superrotation and the magnitude/direction of the resulting hotspot offset are not related in a simple, one-to-one way, but are jointly controlled by rotation, drag, and radiative timescale. The argument above should therefore be regarded as a plausible leading-order mechanism rather than a complete explanation.


Finally, at high rotation rates (third row, right-hand side panels), the hotspot offset remains eastward but is reduced, because rotation inhibits the day-to-nightside transfer of heat. The same reduction is found in the 3D GCM simulations of \citet{Tan2020}, where the eastward offset shrinks as rotation increases.

\begin{figure*}
    \centering
    \includegraphics[width=\linewidth]{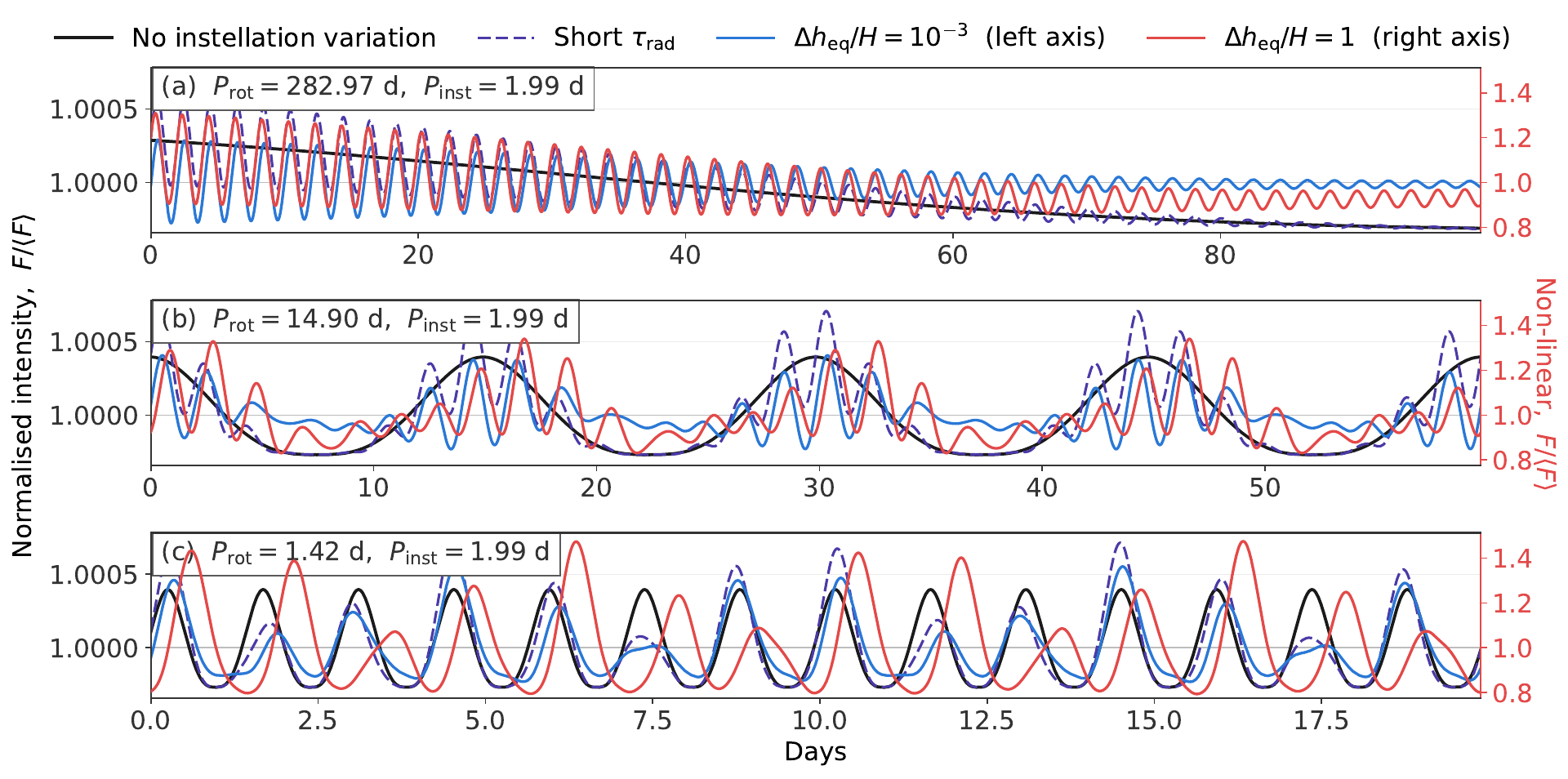}
    \caption{\textbf{Synthetic phase curves exhibit atmospheric variability signatures of tidally locked planets that are variably irradiated.} Synthetic phase curves for the three rotation periods of Figure~\ref{fig:hovmoller}, $P_{\rm rot} = 282.97$, $14.90$ and $1.42$~d in panels (a)--(c), against a fixed instellation period of $1.99$~d. Every phase curve is the disc integral of Equation~\eqref{eq:phase_curve_def} divided by its own time mean, so the axis is the normalized intensity $F/\langle F\rangle$ and each curve is shown at its true amplitude. The forcing equilibrium profile at constant instellation is shown in black, which varies only through the changing viewing geometry. The same forcing with instellation variation included, equivalently the short-$\tau_{\rm rad}$ limit in which the atmosphere follows the forcing instantaneously without producing winds is shown by the purple dashed line. The weakly forced linear response for $\Delta h_{\rm eq}/H = 10^{-3}$ is in blue. These three share the left-hand axis. The strongly forced nonlinear response for $\Delta h_{\rm eq}/H = 1$ (red) is three orders of magnitude larger in amplitude (the red right-hand axis) because of the difference in forcing strengths. The instellation variation here is 50\% ($F=0.5$). The phase curve modulation introduced by instellation variation scales with the forcing strength and is thus larger for strongly forced planets. }
    \label{fig:phase_curve}
\end{figure*}

\begin{figure*}
    \centering
    \includegraphics[width=0.9\linewidth]{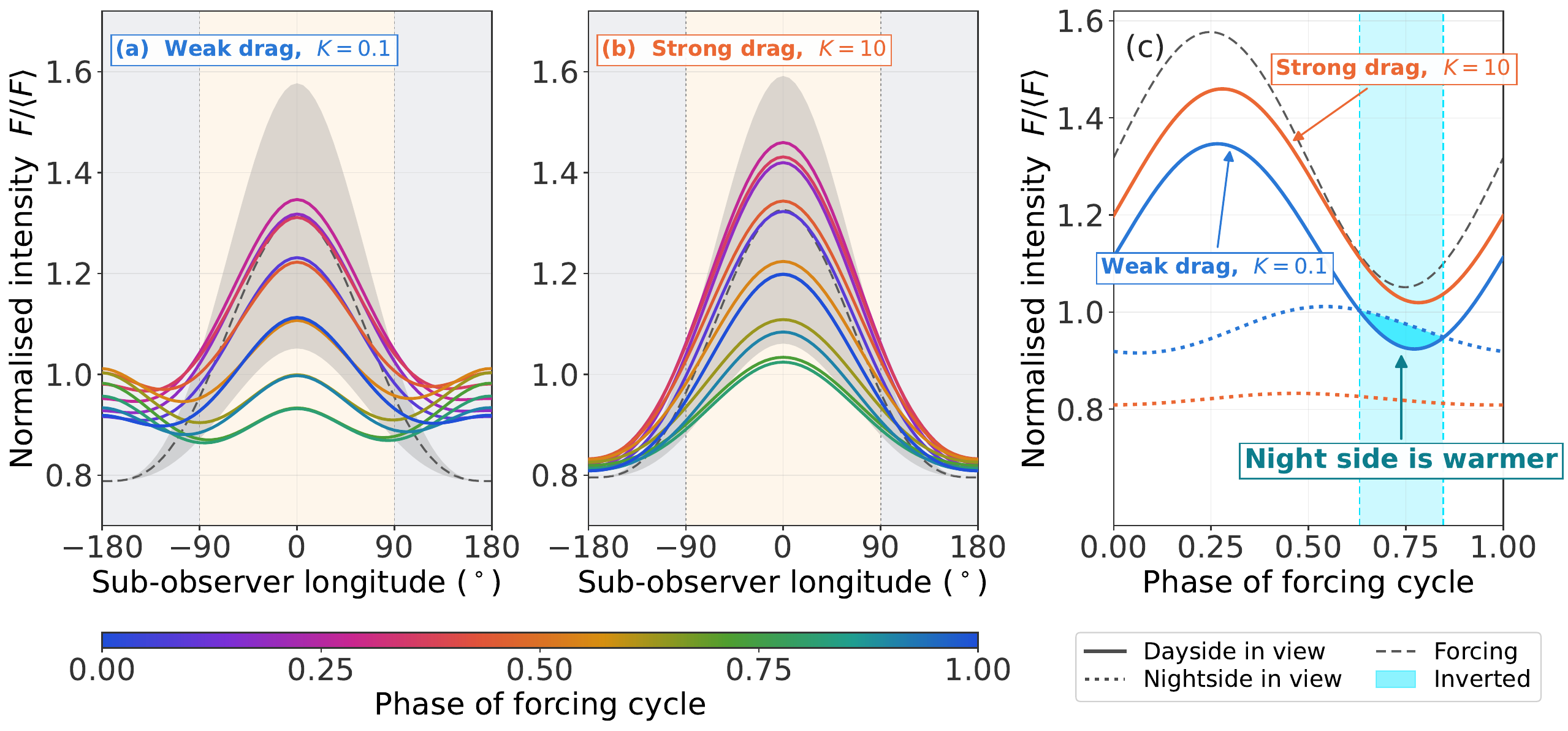}
    \caption{\textbf{Phase curve inversion occurs as a consequence of nightside warming at resonant amplification of wave dominated inertial systems.} Two nonlinear gas-giant runs ($\Delta h_{\rm eq}/H = 1$, $X = 4.745$) that share all parameters except drag, which differs by a factor of $200$. The weak-drag run ($K = 0.1$) sits in the wave-active regime where waves propagate before they are damped, while the strong-drag run sits in the drag-dominated regime. Panels (a) and (b) plot normalized intensity against sub-observer longitude or the longitude above which the observer is directly above the planet, measured with respect to the substellar point, at twelve equally spaced time intervals in one forcing cycle, coloured by phase within that cycle. Shaded background marks the day (yellow) and night (grey) sides, separated by the dotted terminator lines at $\pm90^\circ$. The grey band represents the disc-averaged forcing field over all longitudes and the grey dashed line its mean. Panel (c) illustrates normalized intensity against phase of the forcing cycle, for an observer fixed at the dayside (solid) and at the nightside (dotted). The nightside of the weak-drag run varies four times as much as that of the strong-drag run, and also gets temporarily \textit{higher} in flux than the dayside when the nightside is warmer because of resonant amplification. This nightside warming does not happen in the strong-drag case. }
    \label{fig:phase_curve_drag}
\end{figure*} 1187
1187

The interpretation above rests on a qualitative argument for how wave--mean-flow interactions might suppress the westward, Rossby-dominated pattern found in the linear regime, but we have not yet directly diagnosed the jet or decomposed the flow into its constituent wave modes. There are two directions of investigation that would sharpen this picture, a projection of the solutions onto the free wave modes and a time-dependent Helmholtz decomposition of the flow. First, an explicit projection of both the linear and nonlinear solutions onto the free wave modes of the system, following the methodology of \citet{Shamir2023}, would let us test directly the mechanism of hotspot behaviour in the linear regime. Second, in the nonlinear regime, we may use a time-dependent Helmholtz decomposition of the flow field \citep{hammond2021rotational} to probe the contributions of the divergent, eddy-rotational and eddy-mean components and the timescales over which they grow and decay subject to transient instellation to better understand how the equatorial jet spins up and equilibrates \citep{ShowmanPolvani2011}, which would in turn delineate the hotspot behaviour. We leave the separation of the transient jet dynamics from the quasi-steady wave response to future work. A further regime, in which a much longer radiative timescale admits a freely circulating height pattern whose direction is set by the forcing amplitude rather than by rotation, is presented in Appendix~\ref{app:long-taurad}.

\subsection{Synthetic phase curves}
\label{sec:phase_curves}

The overall climate variability may modulate the observed phase curve of the planet. We now generate synthetic phase curves corresponding to the different cases simulated above. We assume that at a given instant the planet is viewed from a sub-observer longitude $\lambda_{\rm obs}$ in the equatorial plane, so any surface element contributes according to its colatitude $\theta = 90^\circ - \phi$ and its longitude relative to $\lambda_{\rm obs}$ as
$
    \mu = \sin\theta\,\cos(\lambda - \lambda_{\rm obs}),
    \label{eq:mu_def}$
only where $\mu > 0$. The limb is therefore weighted far less than the centre of the disc. The synthetic flux is
\begin{equation}
    F(\lambda_{\rm obs}, t) = \int_{\mu>0} a^2\,h(\lambda,\theta,t)\,\mu\,\sin\theta \,\mathrm{d}\theta\, \mathrm{d}\lambda ,
    \label{eq:phase_curve_def}
\end{equation}
which is the projected screen sketched in the right-hand panel of Figure~\ref{fig:schem_hemi}. This formula is similar to that in Equation (11) of \cite{penn2017thermal}, except the colatitude convention of Dedalus3 switches their cosine factor to sine in our case. The sub-observer longitude is measured from the substellar point, so $0^\circ$ corresponds to the dayside and $\pm180^\circ$ to the nightside. Additionally, in Figure \ref{fig:phase_curve}, each curve is divided by its own time mean, so the vertical axis is the intensity normalized by its mean $F/\langle F\rangle$ which is effectively the fractional variation. The weak forcing cases thus have a much smaller fractional variation (0.0005) compared to the strong forcing ones (0.4). The strong forcing case has been put on the right side axis for convenient visualization in Figure \ref{fig:phase_curve}.

\textbf{Varying rotation.} Figure~\ref{fig:phase_curve} shows phase curves for the slow, intermediate and fast rotation rates of the Hovm\"oller analysis, at a fixed instellation period of $1.99$~d. Both instellation and rotational modes appear distinctly when the two periods are well separated, as in panels (a) and (b), and merge when they are comparable, as in panel (c). In the first two cases the overall variation tracks rotation, while the height of each successive peak is modulated by the instellation available at that moment. The constant instellation case (black) follows the mean of the short radiative time ($\tau_{\rm rad}$, purple dashed). The latter is equivalent to the time varying equilibrium height field $h_{\rm eq}$. The weakly forced linear (blue) and strongly forced nonlinear (red) cases are phase shifted, with the highest peak lagging that of constant instellation; see panel (b). In panel (c), the two periods are similar and hence the modulation is less distinguishable. These plots have been produced at a fractional semi-amplitude $F=0.5$ for the forcing, implying that the instellation received varies by 50\%. The maximum instellation variation for systems presented later in Section~\ref{sec:kw_space} is about $39\%$. Since the response amplitude scales with $F$, we expect the instellation variation to have a significant effect on the modulation of observable phase curves. However, phase curve modulation also scales with the amplitude of planetary response $h_{\rm amp}$. The combined effect for real systems is described in detail later in Section~\ref{sec:kw_space}. Rotation here, at least for the cases shown, does not play any differentiating role per se.


\textbf{Resonant amplification leads to phase curve inversions.} The occurrence of a temporarily warmer nightside has been discussed from theoretical and computational perspectives in the earlier sections. Here in Figure~\ref{fig:phase_curve_drag} we examine the observational effect of such a scenario by varying drag. Here we present multiple phase curves over a single instellation cycle, independent of the rotationally exposed hemisphere of the planet unlike Figure \ref{fig:phase_curve} where the phase curves are a function of rotational/orbital phase. We choose two nonlinear gas-giant runs that share every parameter except the drag timescale: both have $X = 4.745$ and $W = 0.316$, so both lie inside the half-amplitude window of Equation~\eqref{eq:half_amp_window}, (here $0.025 \le W \le 1.76$), and both are forced at $1.50\,W_{\rm nat}$. The distinguishing parameter is $\tau_{\rm drag}$, which differs by a factor of $200$, representing the resonant inertial and damped regimes in Section~\ref{sec:in_dam_resp}. The weak-drag run has is inertial ($K=0.1<X$), allowing resonant amplification of the amplitude. The strong-drag run is damped ($K=10>X$), or damped, allowing amplitude maximization via strong heat retention; see Section~\ref{sec:second_order}. The same effect can also be triggered by rotation, but we choose drag for convenience.

The most notable feature of Figure~\ref{fig:phase_curve_drag} is the consequence of the nightside of the planet becoming temporarily \textit{warmer} than the dayside. This happens for the weak drag case where phase curves do not follow the forcing profiles shown in grey shade, unlike the strong drag case. Resonant amplification and substellar-antistellar hotspot oscillations at weak drag lead to a global heat distribution regime characterized by standing waves, which differs from strongly dragged planets whose response simply follows the forcing. Specifically, at the phase of 0.75 in panel (a), we see that the normalized flux on the nightside (about $\pm 180^{\circ}$, grey shade) is higher than dayside (about $\pm 0^{\circ}$, yellow shade). Panel (c) shows this more explicitly by plotting the day and nightside disc-integrated fluxes individually and the location where the crossover happens for weak drag. Theoretically, this is where the oscillating part exceeds the mean, so the contrast passes through zero and changes sign. For a part of every cycle the disc is brighter on the night hemisphere. In the strong-drag run the crossover never happens. The nightside flux variation is also larger for the weak-drag case (compare the range of flux values at the antistellar point, $\pm 180^{\circ}$, in panels a and b), because the amplified waves carry more heat to the nightside. Thus, the resonance amplified, weak-drag response amplitude has a significantly different signature on phase curves than when the response is maximized by heat retention in the strong-drag limit.

\section{Summary and discussion}\label{sec:summary}

\subsection{Summary of findings}

We have presented an analytical and numerical study of the atmospheric response of tidally locked exoplanets to time-variable stellar irradiation using a periodically forced shallow water model. The central outcome of our reduced-order theoretical development that follows from several simplifying assumptions applied to the fully nonlinear system is that the thermal response of variably irradiated tidally locked planets can be modelled as a forced damped harmonic oscillator. The closed-form analytical planetary response, depicted by day-night height/temperature contrast, is hence oscillatory, with a mean, amplitude and phase-lag. This behaviour is recovered in simulations that follow, quantitatively in agreement with theory (Section~\ref{sec:variable_instellation}).

There are three nondimensional quantities that cover all classes of planets studied with this framework, namely $K=\tau_{\rm wave}^2/(\tau_{\rm rad}\tau_{\rm eff})$, the heat retention parameter that describes the inefficiency of heat redistribution by waves and winds; $W=\omega \tau_{\rm rad}$, the normalized forcing or instellation variation frequency; and $X=\tau_{\rm wave}/\tau_{\rm rad}$, the ratio of gravity wave propagation and radiative timescales of the planet. Gravity waves transport heat by travelling from the dayside to the nightside, while rotation and drag, combined into an effective drag timescale $\tau_{\rm eff}$, prevent such heat redistribution. The time-mean planetary contrast normalized by the forcing strength is found to be $\Delta h_0/\Delta h_{\rm eq} = (1+1/K)^{-1}$, which is a function of just $K$, thus recovering and generalizing the heat redistribution scaling of \citet{perez2013atmospheric} and \citet{komacek2016atmospheric} to the case with variable irradiation (Section~\ref{sec:steady_scaling} and Section~\ref{sec:variable_instellation}). 

There are two regimes of response in the theoretical analysis, the inertial or wave-dominated regime ($K\ll X$) and the damped regime ($K\gg X$). The resonant amplification of response occurs in the inertial regime where waves propagate freely without being damped. Consequently, the natural frequency is the inverse of the wave timescale, $1/\tau_{\rm wave}$ (in the limit $K \to 0$). Even though this study specifically concerns tidally locked planets, we envisage this resonance to have effect for all variably irradiated planets in general. In the damped regime the system effectively reduces to first order. The damped regime also coincides with larger values of $K$, since $K\gg X$, so the amplitude is maximized purely because of the high heat retention capacity. Finally, when $W$ is high, the system response dies out as it is unable to respond to quickly varying irradiation (Figure \ref{fig:foso_theory} and Section~\ref{sec:in_dam_resp}).

Both linear (typically weakly forced with low Rossby number) and nonlinear (strongly forced with high Rossby number) simulations capture the above physics qualitatively, though the former has a better quantitative match. Additionally, simulations unlock global climate patterns and their variability, which is inaccessible via theory (Section~\ref{sec:sim_results}). The principal result concerns the direction of hotspot oscillations, which we find to be controlled primarily by rotation rate rather than drag timescale, even though both have degenerate effects on the thermal response described earlier. Drag-varying experiments at fixed (slow) rotation show purely eastward oscillation, while increasing rotation drives a transition from eastward to westward oscillations. The other parameter controlling the shift is nonlinearity typically triggered by stronger forcing strength ($\Delta h_{\rm eq} \sim H$). The linear, rotation driven east to west transition is suppressed in the nonlinear regime, where nonlinear momentum advection drives all oscillations eastward regardless of rotation rate. This behaviour is consistent with the wave--mean-flow and jet spin-up mechanism of \citet{ShowmanPolvani2010,ShowmanPolvani2011}, \citet{tsai2014three} and \citet{HammondPierrehumbert2018}, in which the equatorial superrotation causes the hotspot to shift eastwards (Section~\ref{sec:discussion}).  

Phase curves trace both rotation and instellation variation when associated periods are significantly different from one another, with instellation variation features scaling linearly with the fractional semi-amplitude of forcing $F$. When resonant amplification occurs, the nightside becomes warmer than the dayside, which appears as a distinctive phase curve inversion because the amplified waves transport more heat to the nightside. The inversion is absent in other cases, where strong rotation or drag damps the waves and the response closely follows the evolving forcing.


\begin{figure*}
    \centering
    \includegraphics[width=\linewidth]{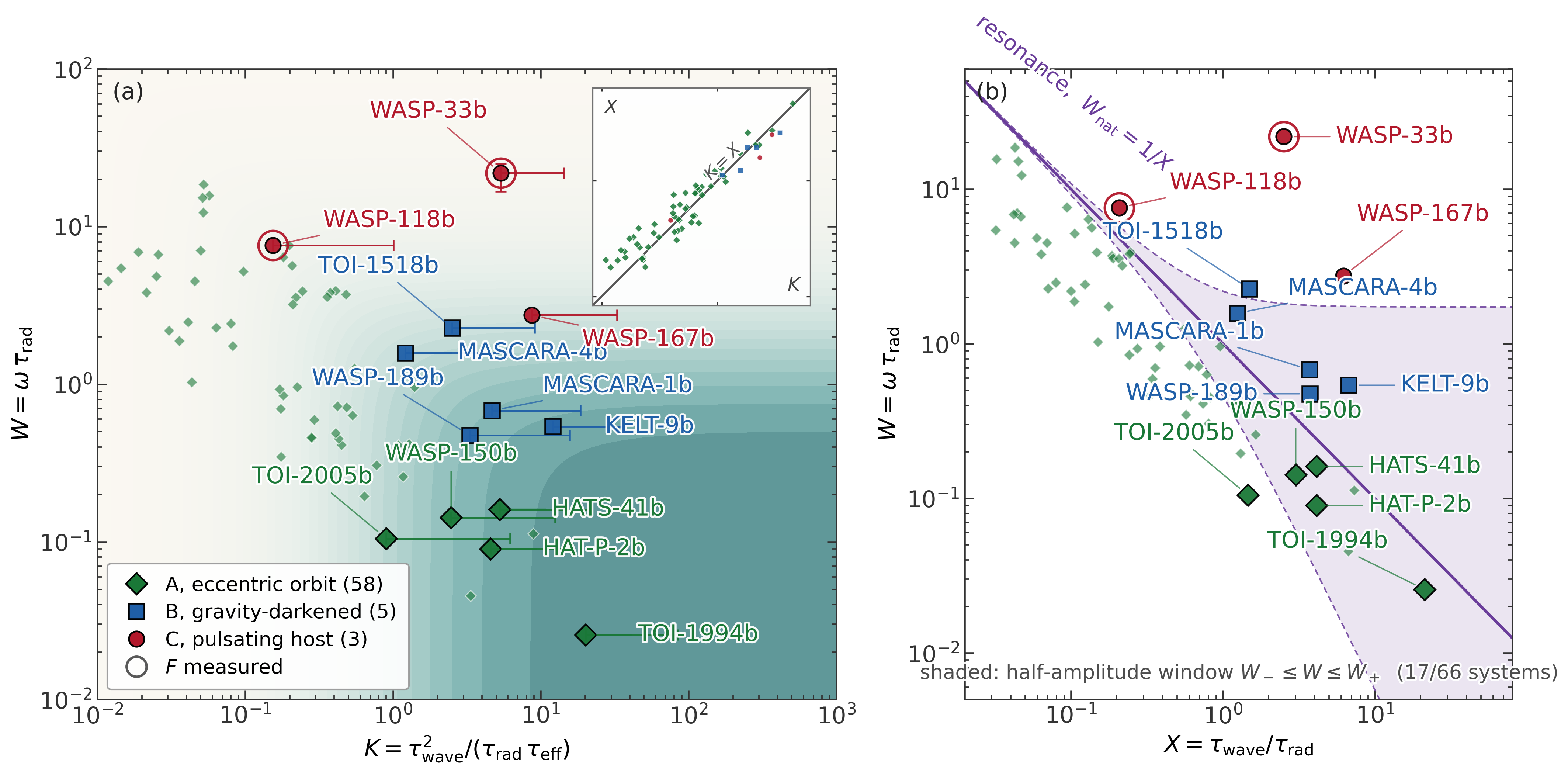}
    \caption{\textbf{At least 10\% net variability is expected on the phase curves of eleven systems because of variable instellation, five of which are near resonance, and HAT-P-2b has the largest variability at 23.28\%; see Table \ref{tab:kw_systems}.} Sixty-six variably irradiated systems placed in the $K$--$W$ plane (a) and the $X$--$W$ plane (b), with $K$, $W$ and $X$ computed for each from archival radii, temperatures and periods (Appendix~\ref{app:kw_calc}). Colour and marker shape encode the variability category, with green diamonds for eccentric orbits (A, 58 systems), blue squares for gravity-darkened hosts (B, 5 systems) and red circles for pulsating hosts (C, 3 systems). Small symbols are unlabelled. Large symbols carry a name and mark the five eccentric planets with the largest predicted modulation $FA$ together with every category-B and C system. An open ring marks the systems whose forcing amplitude is taken from published photometry rather than inferred from the orbit, namely WASP-33b and WASP-118b. Table~\ref{tab:kw_systems} ranks the targets. (a) The shaded field is the first-order amplitude response $\left[(1+1/K)^2+W^2\right]^{-1/2}$, which is independent of $X$ and so applies to the whole sample at once, darker towards the strongly retaining, slowly forced corner. The first-order response carries no restoring force and hence no resonance. Horizontal bars are one-sided and run from the drag-free $K$ at the marker to the strong-drag value at $\tau_{\rm drag} = 10^4$~s. The vertical bar on WASP-33b spans the pulsation spectrum of the host star. (b) Position relative to resonance. The solid purple line is the natural frequency $W_{\rm nat} = 1/X$. The shaded band between the dashed edges is the half-amplitude window $W_- \le W \le W_+$ of Equation~\eqref{eq:half_amp_window}, inside which the $K \to 0$ response exceeds half its peak. The window is symmetric about $W_{\rm nat}$ in $\log W$, has fixed width $W_+ - W_- = \sqrt{3}$, and therefore closes onto the resonance line at small $X$ while opening out at large $X$, where its upper edge tends to $\sqrt{3}$. The window contains seventeen of the sixty-six systems.}
    \label{fig:kw_planets}
\end{figure*}

\subsection{Observational implications for exoplanetary systems}
\label{sec:kw_space}

We now use the above dynamical understanding of variably irradiated tidally locked planets to identify the planets that may have the most discernible effect on observables. Atmospheric variability, such as due to oscillating hotspots described earlier, manifests as additional modulations on planetary phase curves (Figure \ref{fig:phase_curve}), the magnitude of which scales with the fractional semi-amplitude $F$ of the varying forcing. The forcing amplitude $F$ is set by the orbital or stellar parameters of the system and measures how much the instellation received by the planet varies about the mean instellation. In short, planets for which instellation varies significantly are expected to have stronger responses. However, the stellar forcing alone does not ensure a strong planetary response signal. The observed strength of variability in phase curves also scales with the amplitude of oscillation $A=\Delta h_{\rm amp}/F\Delta h_{\rm eq}$ in the response. The response amplitude $A$ depends on the forcing frequency and on planetary parameters such as the heat retention capacity, which is in turn a combination of wave, radiative, rotation and drag timescales; see Equation \eqref{eq:so_amp}. The response amplitude $A$ is maximized under two circumstances, slow instellation variation $W$ paired with strong heat retention $K$ (the damped regime, in the bottom-right corner of the $K$--$W$ space in Figures \ref{fig:foso_theory} and \ref{fig:kw_planets}), and resonant amplification at low $K$ (the inertial regime) around the natural wave frequency of the planet, $\tau_{\rm wave} \sim T_{\rm instel}$ or $W=1/X$ (the dashed white line in Figure \ref{fig:foso_theory} and the line of slope $-1$ in the $X$--$W$ log space of Figure \ref{fig:kw_planets}b). Both $A$ and $F$ need to be large for a strong signal, so the observationally relevant quantity is the product $FA$; see Table \ref{tab:kw_systems}.

We now place observed systems in the $K$--$W$ and $X$--$W$ planes to assess whether their response amplitude should be significant and whether they should lie close to the resonance. Figure~\ref{fig:kw_planets} shows sixty-six systems in three variability categories, namely planets on eccentric orbits, for which the instellation varies over the orbital period between periastron and apoastron (label A, 58 systems); planets on polar orbits around gravity-darkened host stars, for which the instellation varies as the planet passes over the hotter poles and cooler equator of the star (label B, 5 systems); and planets around pulsating host stars (label C, 3 systems). We computed $K$, $X$ and $W$ for each planet from the radii, temperatures and periods in the NASA Exoplanet Archive (Appendix~\ref{app:kw_calc}). The $K$--$W$ and $X$--$W$ parameter spaces differentiate the two different amplitude maximization criteria, namely one for low-frequency, damped maximization (blue region in the left panel), and the second for identifying resonant amplification (purple shading in the right panel). We note that the $K$--$W$ background contour shows only the first-order linearly damped theory and hence does not include the second-order resonance as in Figure \ref{fig:foso_theory}. This is because the resonance lies at a different location for each value of $X$, since $W_{\rm nat}=1/X$, and a single $K$--$W$ space cannot represent all of these locations. The condition of resonance is therefore separately identified in the $X$--$W$ space, where the boundaries of the shaded region include all systems that show resonant amplification of at least 50\% of the maximum value at the exact resonance. The $W_+$ and $W_-$ lines in Figure \ref{fig:foso_theory} are the upper and lower bounds of this region.

\begin{table*}
  \centering
  \caption{Targets ranked by predicted observable modulation $FA$. Of the 66 planets shown in Figure~\ref{fig:kw_planets}, this table reports the twenty largest $FA$ among the eccentric planets (category A), and all gravity-darkened (B) and pulsating-host (C) systems given the small numbers in these two categories. Rows rank $FA$ within each category, so the systems whose forcing amplitude $F$ is unpublished or cannot be inferred fall to the end of the block for that category. The columns give the four timescales that set the nondimensional parameters, namely the wave crossing time $\tau_{\rm wave}$ (Equation~\ref{eq:app_tauwave}), the radiative time $\tau_{\rm rad}$ (Equation~\ref{eq:app_taurad}), the planetary rotation period $P_{\rm rot}$ and the forcing period $T_{\rm instel}$, from which $X = \tau_{\rm wave}/\tau_{\rm rad}$, $W = 2\pi\tau_{\rm rad}/T_{\rm instel}$ and $K$ (Equation~\ref{eq:app_K}, drag-free and hence a lower bound) follow; the normalized amplitude $A \equiv \Delta h_{\rm amp}/(F\Delta h_{\rm eq})$ from Equation~\eqref{eq:so_amp} evaluated at each system's own $(K, W, X)$; the forcing amplitude $F$, computed from the orbital eccentricity via Equation~\eqref{eq:app_F} for category A and taken from the literature for category C; and their product $FA$. Systems in \textbf{bold} are the wave-active resonant ones, meeting both conditions at once, namely that gravity waves cross the planet before they are damped ($K < X$) and that the forcing falls inside the half-amplitude window of resonant amplification in Equation~\eqref{eq:half_amp_window}. In the full sample, eight of the sixty-six systems qualify. The eighth, TOI-3464b ($K = 0.55$, $X = 0.56$), falls outside the twenty largest $FA$ and so has no row here. \label{tab:kw_systems}}
  \small
  \begin{tabular}{llccccccc}
    \toprule
    System & Cat. & $\tau_{\rm wave}$ & $\tau_{\rm rad}$ & $P_{\rm rot}$ &
      $T_{\rm instel}$ & $A$ & $F$ & $FA$ \\
     &  & (d) & (d) & (d) & (d) &  &  &  \\
    \midrule
    HAT-P-2b & A & 0.334 & 0.081 & 1.89 & 5.634 & 0.84 & 0.279 & 0.2328 \\
HATS-41b & A & 0.443 & 0.107 & 2.17 & 4.194 & 0.87 & 0.197 & 0.1713 \\
\textbf{TOI-1994b} & \textbf{A} & \textbf{0.351} & \textbf{0.017} & \textbf{2.33} & \textbf{4.034} & \textbf{0.96} & \textbf{0.176} & \textbf{0.1694} \\
TOI-2005b & A & 0.421 & 0.289 & 4.30 & 17.31 & 0.49 & 0.331 & 0.1612 \\
\textbf{WASP-150b} & \textbf{A} & \textbf{0.385} & \textbf{0.128} & \textbf{2.95} & \textbf{5.644} & \textbf{0.76} & \textbf{0.196} & \textbf{0.1487} \\
CoRoT-20b & A & 0.361 & 0.450 & 2.36 & 9.243 & 0.45 & 0.326 & 0.1473 \\
HAT-P-34b & A & 0.423 & 0.361 & 2.42 & 5.453 & 0.60 & 0.227 & 0.1372 \\
\textbf{TOI-4603b} & \textbf{A} & \textbf{0.350} & \textbf{0.053} & \textbf{4.36} & \textbf{7.246} & \textbf{0.82} & \textbf{0.167} & \textbf{0.1367} \\
XO-3b & A & 0.418 & 0.057 & 2.16 & 3.192 & 0.93 & 0.142 & 0.1320 \\
\textbf{TOI-5301b} & \textbf{A} & \textbf{0.396} & \textbf{0.241} & \textbf{3.48} & \textbf{5.859} & \textbf{0.61} & \textbf{0.170} & \textbf{0.1039} \\
\textbf{WASP-186b} & \textbf{A} & \textbf{0.411} & \textbf{0.328} & \textbf{2.99} & \textbf{5.027} & \textbf{0.59} & \textbf{0.170} & \textbf{0.1010} \\
HATS-40b & A & 0.474 & 0.499 & 2.03 & 3.264 & 0.61 & 0.160 & 0.0972 \\
HD~17156b & A & 0.505 & 1.544 & 3.67 & 21.22 & 0.23 & 0.390 & 0.0880 \\
TOI-172b & A & 0.384 & 0.294 & 4.90 & 9.477 & 0.44 & 0.198 & 0.0870 \\
TIC~393818343b & A & 0.527 & 1.534 & 3.89 & 16.25 & 0.24 & 0.337 & 0.0796 \\
WASP-162b & A & 0.456 & 0.751 & 4.25 & 9.625 & 0.33 & 0.228 & 0.0743 \\
HATS-27b & A & 0.507 & 2.738 & 1.23 & 4.637 & 0.23 & 0.320 & 0.0737 \\
TOI-2025b & A & 0.428 & 0.581 & 4.42 & 8.872 & 0.35 & 0.205 & 0.0725 \\
HATS-10b & A & 0.356 & 1.887 & 1.18 & 3.313 & 0.23 & 0.269 & 0.0610 \\
HD~118203b & A & 0.425 & 0.709 & 3.81 & 6.135 & 0.37 & 0.161 & 0.0588 \\
KELT-9b & B & 0.425 & 0.064 & 1.48 & 0.7406 & 0.88 & \nodata & \nodata \\
MASCARA-1b & B & 0.432 & 0.116 & 2.15 & 1.074 & 0.83 & \nodata & \nodata \\
\textbf{MASCARA-4b} & \textbf{B} & \textbf{0.440} & \textbf{0.354} & \textbf{2.82} & \textbf{1.412} & \textbf{0.59} & \nodata & \nodata \\
TOI-1518b & B & 0.512 & 0.344 & 1.90 & 0.9513 & 0.42 & \nodata & \nodata \\
\textbf{WASP-189b} & \textbf{B} & \textbf{0.385} & \textbf{0.103} & \textbf{2.72} & \textbf{1.362} & \textbf{0.89} & \nodata & \nodata \\
WASP-33b & C & 0.416 & 0.166 & 1.22 & 0.0476 & 0.05 & 0.001 & $4.6\times10^{-5}$ \\
WASP-118b & C & 0.477 & 2.298 & 4.05 & 1.9 & 0.18 & $2.0\times10^{-4}$ & $3.6\times10^{-5}$ \\
WASP-167b & C & 0.451 & 0.073 & 2.02 & 0.1667 & 0.34 & \nodata & \nodata \\
\bottomrule
  \end{tabular}
\end{table*}

We also plot all planets in $K$--$X$, a complementary space to identify the separation in inertial and damped responses; see the inset at the top-right corner of the $K$--$W$ panel. We find that every system lies close to the $K = X$ boundary between the inertial and damped regimes described in Section~\ref{sec:in_dam_resp}. In the drag-free limit the ratio is simply $K/X = \tau_{\rm wave}\Omega$, and it spans only $0.28$--$2.59$ across the whole sample, with a median of $0.87$. The clustering is a generic property of close-in giants and does not come from the selection of the sample. Tidally locked and pseudo-synchronous giants have $\tau_{\rm wave} \sim 0.3$--$0.6$~d and rotation periods of a few days, so $\tau_{\rm wave}\Omega$ cannot be far from unity. Observed close-in giants therefore populate the transition between the two dynamical regimes. 

The resonant amplification predicted in Section~\ref{sec:variable_instellation} may be observationally accessible. The half-amplitude window of Equation~\eqref{eq:half_amp_window}, within which the normalized amplitude $A$ takes a value of at least 0.5, hosts multiple systems. From the $X$--$W$ plot, we find that seventeen of the sixty-six systems fall inside the window, namely thirteen of the fifty-eight eccentric planets, four of the five gravity-darkened hosts and none of the three pulsators. The gravity-darkened systems are the category most reliably close to experiencing resonant amplification. We note that the window is derived in the $K \to 0$ limit, whereas the sample has $K \sim X$, so the window should be read only as an indicator of the resonance. We need to additionally check whether $K<X$ to make sure the amplification is resonant. If $K>X$, the amplification follows the response of a first-order system, which lacks the occasional nightside warming of the resonant response and therefore has a different phase curve signature; see Figure \ref{fig:phase_curve_drag}. Of the seventeen systems identified within the resonant boundaries, we find eight systems to satisfy $K\lesssim X$, seven of which feature in bold in Table \ref{tab:kw_systems}.  


Finally, we rank in Table~\ref{tab:kw_systems} a subset of these planets, namely the twenty eccentric planets with the largest predicted modulation together with all planets in categories B and C. The ranking uses the observable modulation $FA$, the product of the forcing amplitude and the response amplitude, which quantifies the strength of the variability superimposed on the constant-instellation response. The eccentric planets dominate the list, and every eccentric planet in Table~\ref{tab:kw_systems} has a larger $FA$ than any pulsating system, with HAT-P-2b first at $FA = 0.23$. This is because an eccentric orbit modulates the instellation far more than stellar pulsation, which supplies only millimagnitudes of variation. The modulation from gravity darkening has not been measured for any planet in the sample, so the gravity-darkened planets are not ranked. 

Among the twenty eccentric planets in Table~\ref{tab:kw_systems}, ten also fall inside the half-amplitude window, and five of the ten have $K<X$, so resonance amplifies the response of these five planets in addition to the strong forcing. HAT-P-2b is the strongest predicted signal in the sample ($FA = 0.23$, $A = 0.84$) and lies inside the half-amplitude window, even though $K>X$ for HAT-P-2b. HATS-40b ($FA = 0.097$) sits essentially \emph{on} the resonance line, at $0.91\,W_{\rm nat}$, but has $K>X$, so the large response of HATS-40b comes from heat retention in the damped regime rather than from resonant amplification. The best candidates for studying resonant amplification are therefore the five eccentric planets in bold in Table~\ref{tab:kw_systems}, led by TOI-1994b ($FA = 0.17$) and WASP-150b ($FA = 0.15$). We note the expected difference between the signatures of resonant amplification and of damped, low-frequency amplitude maximization. Resonant amplification necessarily includes hotspot oscillations, while damped, low-frequency maximization does not, as seen in Figures \ref{fig:earth_analog} and \ref{fig:hot_Jupiter}. The resonant hotspot motion leads to temporary nightside warming, which may be observed in phase curves as shown in Section~\ref{sec:phase_curves} (Figure~\ref{fig:phase_curve_drag}). The nightside warming is absent for damped, low-frequency amplification.

In conclusion, eccentric, pseudo-synchronous giants are the targets of choice for detecting variable-irradiation signatures on their phase curves and spectra. The eccentric planets that combine large forcing amplitude $F$ with a near-resonant forcing frequency $W$ outrank those with larger $F$ alone, because of resonant amplification. Among the non-eccentric systems, only the gravity-darkened planets have observational prospects. The window contains four of the five gravity-darkened planets, with strong planetary response amplitudes $A$ between $0.59$ and $0.89$, so the predicted response of these planets is as large as that of any eccentric planet. However, the forcing amplitude of the gravity-darkened planets is unmeasured, which prevents a ranking of the gravity-darkened planets. Pulsating hosts, the category that originally motivated this study, are far above resonance, and the forcing amplitudes of pulsating hosts are two to three orders of magnitude smaller than those of the eccentric planets, which is too small to have a distinguishable impact on phase curve variability. A more careful study is nonetheless required to ascertain $FA$ for the targets, in particular for the eccentric planets, which are pseudo-synchronous rather than perfectly tidally locked, so that the rotation period of each eccentric planet differs from the orbital period that sets the instellation variation period.

\subsection{Future work}

Several avenues remain for future work. The shallow water model neglects vertical structure, baroclinic instability, magnetic drag, and cloud feedbacks, all of which can modify phase curve morphology. 3D GCM simulations will be needed to test whether the resonance and oscillation-direction predictions survive in more realistic atmospheres. The mechanism underlying the westward oscillations in the linear, fast-rotation regime and their nonlinear suppression warrants further study in terms of angular momentum budget and wave--mean flow interactions. We envisage the resonance effect triggered by a match between the wave and forcing frequencies to occur also for asynchronous planets, so the effect of variable irradiation would extend to a wider range of systems and to more observational targets. Specifically, variability signals for asynchronous planets could be probes into their rotation rate and other planetary features.

\section*{Acknowledgements}

D.B. would like to thank Kristen Menou and the University of Toronto for providing financial support to carry out this work. This work benefited from the 2025 Exoplanet Summer Program in the Other Worlds Laboratory (OWL) at the University of California, Santa Cruz, a programme partially supported by funding from NASA. A significant portion of the theoretical development was carried out there. D. B. would also like to thank Heather Knutson for constructive discussions. 

This work made use of \textsc{numpy} \citep{harris2020array}, \textsc{scipy} \citep{virtanen2020scipy} and \textsc{dedalus} \citep{burns2020dedalus}, and of the large language model assistants Claude, Gemini and ChatGPT.


\section*{Data Availability}

The shallow water model used in this work is built on the open-source spectral
framework \textsc{dedalus}~v3 \citep{burns2020dedalus}. The model source code,
the run scripts for some of the simulation suites described in
Section~\ref{sec:simulations}, and the post-processing notebooks are publicly available at
\url{https://github.com/Textydeep/variably-irradiated-tidally-locked-planets.git}. The
planetary and stellar parameters in Table~\ref{tab:kw_systems} and
Figure~\ref{fig:kw_planets} were taken from the NASA Exoplanet Archive
(\url{https://exoplanetarchive.ipac.caltech.edu}).

\bibliographystyle{mnras}
\bibliography{tlp_mnras}

\appendix

\section{Which linearity condition breaks first, and what happens when only one does}
\label{sec:app_linearity}

In Section~\ref{sec:theory} we required both the mass condition \eqref{eq:lin_condt2} and the
momentum condition \eqref{eq:lin_condt1} to hold for the linear theory to be valid. Here we
show that the two conditions are not independent, identify which condition is the binding one
in a given part of parameter space, and describe how the flow behaves when only one condition
is violated. We write
$\delta_{\rm eq} = \Delta h_{\rm eq}/H$ for the nondimensional forcing and use Equations \eqref{eq:lin_condt2} and \eqref{eq:lin_condt1} to redefine
\begin{equation}
    \varepsilon_h \;\equiv\; \frac{\Delta h}{H} \;=\; \frac{\delta_{\rm eq}}{1+1/K},
    \quad \text{and} \quad
    \varepsilon_u \;\equiv\; \frac{U\tau_{\rm eff}}{a}
    \;=\; \frac{\tau_{\rm eff}^{2}}{\tau_{\rm wave}^{2}}\,\frac{\delta_{\rm eq}}{1+1/K}.
    \label{eq:app_eps}
\end{equation}
Linearity requires $\varepsilon_h \ll 1$ and $\varepsilon_u \ll 1$. The quantity
$\varepsilon_u$ is the drag-modified Rossby number of the flow, as mentioned in the main text. When we divide the momentum condition by the mass condition,
and use $K = \tau_{\rm wave}^2/(\tau_{\rm rad}\tau_{\rm eff})$ together with
$X = \tau_{\rm wave}/\tau_{\rm rad}$, we obtain
\begin{equation}
    \frac{\varepsilon_u}{\varepsilon_h}
    \;=\; \left(\frac{\tau_{\rm eff}}{\tau_{\rm wave}}\right)^{\!2}
    \;=\; \left(\frac{X}{K}\right)^{\!2},
    \quad\text{or equivalently}\quad
    \varepsilon_u \;=\; \frac{X^{2}\,\delta_{\rm eq}}{K^{2}+K}.
    \label{eq:app_ratio}
\end{equation}
The two conditions therefore differ only by the squared ratio of the damping timescale to the
wave timescale, and the two conditions become equal when $K = X$. The line $K = X$ is the same line that separates the
wave-active and drag-dominated regimes in Section~\ref{sec:in_dam_resp}. We may therefore divide the parameter
space into two cases, $K < X$ and $K > X$. For $K < X$, which corresponds to weak drag or slow rotation
($\tau_{\rm eff} > \tau_{\rm wave}$), the momentum condition is the binding one and is violated while
$\Delta h/H$ is still small. For $K > X$, which corresponds to strong drag or fast rotation
($\tau_{\rm eff} < \tau_{\rm wave}$), the mass condition is the binding one and is violated while the
winds are still slow.
Only in the limit $K \gg X$ does the single criterion $\delta_{\rm eq} \ll 1$ used by
\citet{perez2013atmospheric} become sufficient on its own. We note that the
Froude number of the flow, $\mathrm{Fr} = U/c$, satisfies $\mathrm{Fr}^{2} =
\varepsilon_h\varepsilon_u$ when evaluated with the linear scalings, so the two conditions
jointly control how close the flow is to the gravity wave speed.

\subsection{Case I, nonlinear momentum and linear mass ($K < X$)}
\label{sec:app_caseI}

When only $\varepsilon_u$ exceeds unity, the height perturbation is still a small fraction of
the layer depth, so the mass balance \eqref{eq:scale2} remains linear, but momentum advection
is no longer negligible against the effective drag. We may then keep the advection term in the
momentum balance while retaining the linear mass balance,
\begin{equation}
    \frac{g\,\Delta h}{a} \;\sim\; \frac{U}{\tau_{\rm eff}} + \frac{U^{2}}{a},
    \quad
    \frac{HU}{a} \;\sim\; \frac{\Delta h_{\rm eq} - \Delta h}{\tau_{\rm rad}} .
    \label{eq:app_caseI_bal}
\end{equation}
The mass balance gives the velocity directly as $U \sim a\,s/\tau_{\rm rad}$, where
$s = \delta_{\rm eq} - \delta$ and $\delta = \Delta h/H$. When we substitute this velocity into the momentum
balance, we obtain a quadratic relation in place of Equation \eqref{eq:scale3},
\begin{equation}
    \delta \;=\; K s + X^{2} s^{2},
    \quad
    s \;=\; \frac{-(1+K) + \sqrt{(1+K)^{2} + 4X^{2}\delta_{\rm eq}}}{2X^{2}} .
    \label{eq:app_caseI_sol}
\end{equation}
When we set $X \to 0$, the advection term vanishes and we recover Equation \eqref{eq:scale3} exactly. For small
$K$, Equation \eqref{eq:app_caseI_sol} reduces to
\begin{equation}
    \frac{\Delta h}{H} \;\simeq\; \frac{\Delta h_{\rm eq}}{H}
    \left[\,K + X^{2}\,\frac{\Delta h_{\rm eq}}{H}\,\right],
    \label{eq:app_caseI_lowK}
\end{equation}
in which the ratio of the second term to the first is precisely $\varepsilon_u$. The advective
contribution therefore takes over exactly when the momentum condition fails, as expected.

The physical consequence is that the day--night contrast does not vanish as the damping is removed. The linear result \eqref{eq:scale3} predicts $\Delta h \to 0$ as $K \to 0$, because with no drag and no rotation there is nothing to balance the pressure gradient. Momentum advection supplies that balance instead, and sets a floor $\Delta h/H \to X^{2}(\Delta h_{\rm eq}/H)^{2}$ that is independent of $K$. In a plot of $\Delta h/H$ against $K$, the linear prediction continues to fall towards small $K$ while the nonlinear solution flattens out, which is the behaviour of the nonlinear simulations. The flattening is the shallow-water counterpart of the advective limit of \citet{komacek2016atmospheric}, in which the drag rate is supplemented by $U/a$.

\begin{figure*}
  \centering
  \includegraphics[width=0.9\textwidth]{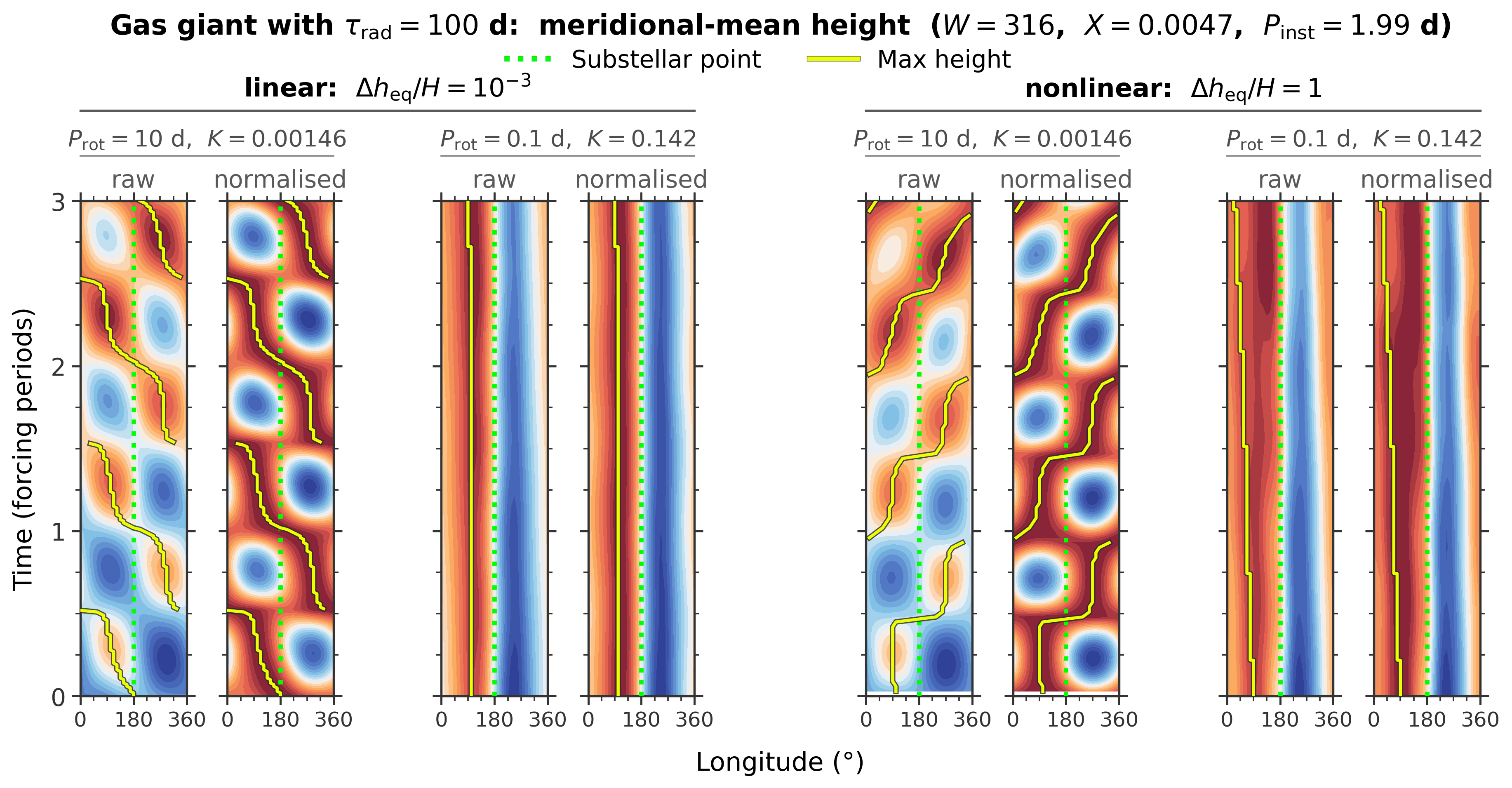}
  \caption{Hovm\"oller diagrams of the meridional-mean height (averaged over
  $\pm45^\circ$ latitude) for the $\tau_{\rm rad} = 100\,$d gas giant, over the last
  three forcing periods. Each case is shown raw and normalized by its
  instantaneous maximum. The left group has linear forcing, $\Delta h_{\rm eq}/H = 10^{-3}$,
  and the right group nonlinear forcing, $\Delta h_{\rm eq}/H = 1$. Within each group the rotation
  rate increases from left to right, $P_{\rm rot} = 10\,$d ($K = 1.5\times10^{-3}$)
  and $0.1\,$d ($K = 0.14$). Dotted green marks the substellar meridian, yellow
  the longitude of maximum height. All four share $W = 316$, $X = 4.7\times10^{-3}$
  and $T_{\rm instel} = 1.99\,$d.}
  \label{fig:hov-taurad100}
\end{figure*}

The wind speed itself is largely unaffected. When $\Delta h \ll \Delta h_{\rm eq}$ the mass
balance alone fixes $U \sim a\,\delta_{\rm eq}/\tau_{\rm rad}$, independently of which terms
balance the pressure gradient, so it is the height contrast and not the velocity that responds
to the loss of linearity here. Beyond these scalings, Case I is also the regime in which eddy
momentum fluxes can spin up an equatorial jet \citep{ShowmanPolvani2011}, which then Doppler
shifts the forced wave pattern and displaces the hotspot eastward
\citep{HammondPierrehumbert2018, tsai2014three}. Neither effect is contained in the linear
theory. Since $\mathrm{Fr}^{2} = \varepsilon_h \varepsilon_u$, the flow can also approach the
gravity wave speed in this case even though $\Delta h/H$ remains small.

\subsection{Case II, nonlinear mass and linear momentum ($K > X$)}
\label{sec:app_caseII}

When only $\varepsilon_h$ exceeds unity, the winds remain slow enough that the momentum
equation stays in the balanced form in which the velocity is diagnosed from the
height field through $\mathbf{u}' = -g\tau_{\rm eff}\nabla h'$. Case II is the more benign of the
two cases, because that diagnostic relation may be substituted into the \emph{full} mass
Equation \eqref{eq:SW_mass} without any expansion in $h'/H$. The substitution gives a single closed
equation for the height,
\begin{equation}
    \frac{\partial h}{\partial t}
    - \frac{g\,\tau_{\rm eff}}{2}\,\nabla^{2}\!\left(h^{2}\right)
    + \frac{h}{\tau_{\rm rad}}
    \;=\; \frac{h_{\rm eq}(\lambda,\phi,t)}{\tau_{\rm rad}},
    \label{eq:app_caseII_pde}
\end{equation}
which is the nonlinear counterpart of a linearized first-order theory derived from a balanced flow (not shown here), where the constant diffusivity
$gH\tau_{\rm eff}$ is replaced by the state-dependent diffusivity $g h \tau_{\rm eff}$. Equation \eqref{eq:app_caseII_pde} is still first order in time. The qualitative
conclusions of the first-order theory therefore survive, and the response to variable instellation
remains a damped relaxation with a monotonically decreasing amplitude and a phase lag bounded
by $\pi/2$, and no resonance appears. A violation of the mass condition alone thus does not modify the structure of the response.

Hence, for $K < X$ the momentum condition binds, and the violation of the momentum condition sets a floor on the day--night contrast without changing the wind speed. For $K > X$ the mass condition binds, and the violation of the mass condition leaves the damped, non-resonant structure of the response intact. Appendix~\ref{app:long-taurad} examines a region of parameter space that the simulations of Section~\ref{sec:simulations} do not sample, the limit of long radiative timescale, where the forcing oscillates far faster than the atmosphere can relax.

\section{Response at long radiative timescale}
\label{app:long-taurad}

The two planets of Section~\ref{sec:simulations} have $X \equiv \tau_{\rm wave}/\tau_{\rm rad}$
of order unity. Because $K \propto \tau_{\rm rad}^{-1}$ and $W \propto \tau_{\rm rad}$,
a much longer radiative timescale moves the system into a corner of the $K$--$W$
plane that those sweeps do not sample. To probe this corner, we repeat the gas giant
($g = 10\,{\rm m\,s^{-2}}$, $a = 8.2\times10^{7}\,{\rm m}$, $H = 4\times10^{5}\,{\rm m}$,
$\tau_{\rm drag} = 3.89\times10^{6}\,{\rm s}$, $T_{\rm instel} = 1.99\,{\rm d}$)
with $\tau_{\rm rad} = 100\,$d instead of $0.1\,$d, while we hold the planet, the drag and
the orbit fixed. The longer radiative timescale gives $X = 4.7\times10^{-3}$ and $W = 316$, so the forcing
oscillates far faster than the atmosphere can relax. We run four cases spanning
the linear and nonlinear regimes ($\Delta h_{\rm eq}/H = 10^{-3}$ and $1$) and slow
and fast rotation ($P_{\rm rot} = 10\,$d and $0.1\,$d, i.e., $K = 1.5\times10^{-3}$
and $0.14$), integrating each for 40 forcing periods at the resolution of
Section~\ref{sec:simulations} and diagnosing the last three.\footnote{The timestep must resolve
the rotation period as well as the forcing period, so we take
$\Delta t = \min(T_{\rm instel}/100,\ 0.08\,{\rm d},\ P_{\rm rot}/100)$.}

Figure~\ref{fig:hov-taurad100} shows the outcome. Since $K \ll 1$ throughout, the
day--night contrast of the meridional-mean height is strongly suppressed,
and reaches at most $4.3\%$ of $\Delta h_{\rm eq}$, so the normalized panels are therefore
the informative ones. There are two behaviours, a circulating pattern at slow rotation and a stationary pattern at fast rotation. At
$P_{\rm rot} = 10\,$d the height pattern circulates steadily instead of locking to the substellar
meridian, and completes one circuit per forcing period
($2.04\,$d and $2.07\,$d against $T_{\rm instel} = 1.99\,$d). The circulation is westward under weak
forcing and \emph{eastward} under strong forcing, the same sense reversal
reported in Section~\ref{sec:hovmoller} at $K = 1$. At $P_{\rm rot} = 0.1\,$d the
pattern is instead stationary (the zonal wavenumber-1 crest drifts by less than
$1^\circ\,{\rm day}^{-1}$), locked $82^\circ$ west of substellar in the linear case
and $140^\circ$ west in the nonlinear one. Hence, long radiative timescales weaken the
response, and the height pattern may also circulate freely in a direction set by the forcing
amplitude.

\section{Computation of the nondimensional parameters for observed systems}
\label{app:kw_calc}

This appendix describes how $K$, $W$, $X$ and the forcing amplitude $F$ in Figure~\ref{fig:kw_planets} and Table~\ref{tab:kw_systems} are obtained. Planetary radii $R_{\rm p}$, masses, equilibrium temperatures $T_{\rm eq}$, orbital periods $P_{\rm orb}$ and eccentricities $e$ are the archival values from the NASA Exoplanet Archive composite-parameters table \citep{NEA2026}, adopted as published, without any rescaling of $T_{\rm eq}$ to the dayside or to periastron. The surface gravity follows from the archival mass and radius.

\subsection{Active-layer depth and the wave timescale}

The shallow-water layer depth $H$ is an equivalent depth rather than a measured quantity \citep{vallis2017atmospheric}. We set $H$ to the pressure scale height evaluated at the equilibrium temperature of each planet,
\begin{equation}
    H = \frac{R_{\rm sp} T_{\rm eq}}{g}, \quad R_{\rm sp} = \frac{\mathcal{R}}{\mu},
    \label{eq:app_H}
\end{equation}
with $\mu = 2.3$ for a solar-composition H/He atmosphere. For the generic hot Jupiter of \cite{perez2013atmospheric}, with $g = 10\,\mathrm{m\,s^{-2}}$ and $T_{\rm eq} = 1600$~K, Equation \eqref{eq:app_H} returns $H = 578$~km, which is of the same order as the value of $400$~km adopted by \cite{perez2013atmospheric}. The resulting depths across the sample range from $21$~km for the high-gravity TOI-4603b to ${\sim}1600$~km for the inflated TOI-615b.

The gravity-wave speed is
\begin{equation}
    c = \sqrt{gH} = \sqrt{R_{\rm sp} T_{\rm eq}},
\end{equation}
which is independent of $g$, so that the wave timescale
\begin{equation}
    \tau_{\rm wave} = \frac{R_{\rm p}}{\sqrt{R_{\rm sp} T_{\rm eq}}}
    \label{eq:app_tauwave}
\end{equation}
depends only on the planetary radius and temperature. 

The radiative timescale is the relaxation time of the weather layer cooling as a blackbody at its own equilibrium temperature, as in Appendix A of \cite{Banik2025},
\begin{equation}
    \tau_{\rm rad} = \frac{\Delta p}{g}\,\frac{c_{p}}{4\sigma T_{\rm eq}^{3}},
    \label{eq:app_taurad}
\end{equation}
with $\Delta p$ the thickness of the weather layer, $c_{p}$ the specific heat at constant pressure, and $\sigma$ the Stefan--Boltzmann constant. We take $\Delta p \simeq 10^{5}$~Pa, the photosphere depth appropriate to a hydrogen-dominated giant, and $c_{p} = 14.304$~kJ~K$^{-1}$~kg$^{-1}$, both following \cite{Banik2025}. $X = \tau_{\rm wave}/\tau_{\rm rad}$ follows directly. The radiative timescale carries a factor $1/g$ and therefore depends on the planetary mass, so $X$, $K$ and $W$ also depend on the planetary mass.

\subsection{Rotation, drag and $K$}

Equation~\eqref{eq:tau_eff} gives $\tau_{\rm eff}^{-1} = 2\Omega\sin\phi_0 + \tau_{\rm drag}^{-1}$.
We evaluate the effective drag at a reference latitude $\phi_0 = 30^\circ$, midway between the equator and the
pole, so that the Coriolis parameter is $f = 2\Omega\sin 30^\circ = \Omega$ and
$\tau_{\rm eff}^{-1} = \Omega + \tau_{\rm drag}^{-1}$. The heat-retention parameter is then
\begin{equation}
    K = \frac{\tau_{\rm wave}^{2}}{\tau_{\rm rad}\tau_{\rm eff}}
      = X\,\tau_{\rm wave}\left(\Omega + \tau_{\rm drag}^{-1}\right).
    \label{eq:app_K}
\end{equation}
The drag timescale is unconstrained for every system in the sample. We therefore quote $K$ in the drag-free limit $\tau_{\rm drag}\to\infty$, where $K = X\,\tau_{\rm wave}\Omega$. Because drag can only shorten $\tau_{\rm eff}$, the drag-free $K$ is a strict lower bound, and Figure~\ref{fig:kw_planets} shows the one-sided departure to a strong-drag case for which we set $\tau_{\rm drag} = 10^{4}$~s.

Rotation rates are assigned as follows. For circular orbits the planet is taken to rotate synchronously, $\Omega = 2\pi/P_{\rm orb}$. For eccentric orbits we use the pseudo-synchronous spin rate of \cite{Hut1981},
\begin{equation}
    \frac{\Omega_{\rm ps}}{n} = \frac{1 + \tfrac{15}{2}e^{2} + \tfrac{45}{8}e^{4} + \tfrac{5}{16}e^{6}}
                                     {\left(1 + 3e^{2} + \tfrac{3}{8}e^{4}\right)\left(1-e^{2}\right)^{3/2}},
    \label{eq:app_hut}
\end{equation}
with $n = 2\pi/P_{\rm orb}$. 

\subsection{Forcing period and amplitude}

The forcing period $T_{\rm instel}$ is $P_{\rm orb}$ for the eccentric planets, the dominant pulsation period for the pulsating hosts, and $P_{\rm orb}/2$ for the gravity-darkened hosts, because a planet on a misaligned orbit passes over the hotter stellar poles twice per orbit \citep{Ahlers2020}. The dimensionless frequency is then $W = 2\pi\tau_{\rm rad}/T_{\rm instel}$.

For the eccentric planets the instellation scales as $r^{-2}$, so $T_{\rm eq}\propto r^{-1/2}$, and we take the layer thickness to be proportional to $T_{\rm eq}$. The difference between the thickness at periastron, $r = a(1-e)$, and at apoastron, $r = a(1+e)$, divided by the sum of the two thicknesses, gives the fractional semi-amplitude
\begin{equation}
    F = \frac{\sqrt{\rho}-1}{\sqrt{\rho}+1}, \quad
    \rho = \frac{1+e}{1-e},
    \label{eq:app_F}
\end{equation}
with $F = 0.13$--$0.39$ across the category-A sample.

For WASP-33 and WASP-118, $F$ is the pulsation amplitude, ${\sim}10^{-3}$ for the millimagnitude pulsations of WASP-33 \citep{vonEssen2020} and ${\sim}200$~ppm for WASP-118 \citep{Mocnik2017}. No published measurement of the modulation in received flux exists for any gravity-darkened planet in the sample, so all five category-B systems, together with WASP-167b in category C, are placed on the diagrams but omitted from the $FA$ ranking.

\subsection{Sample selection}

Category B comprises the planets on strongly misaligned orbits around the rapidly rotating, gravity-darkened stars KELT-9, MASCARA-1, MASCARA-4, TOI-1518 and WASP-189 \citep{Ahlers2020, Hooton2022, Ahlers2020m4, Cabot2021, Lendl2020}. Category C comprises the planets of the pulsating hosts WASP-33, WASP-118 and WASP-167 \citep{Herrero2011, Mocnik2017, Temple2017}. 
The category-A sample is drawn from \cite{NEA2026} for transiting planets
\begin{equation}
    e \geq 0.25,\quad R_{\rm p} \geq 0.8\,R_{\rm J},\quad T_{\rm eq} \geq 800~\mathrm{K},\quad
    3~\mathrm{d} \leq P_{\rm orb} \leq 25~\mathrm{d}.
\end{equation}
The radius and temperature cuts keep the sample within the giant-planet regime where the shallow-water framework and Equation~\eqref{eq:app_taurad} apply. The lower period bound removes ultra-short-period orbits whose reported eccentricities are unlikely to survive tidal circularization, and the upper bound keeps the sample where pseudo-synchronous rotation is plausible. The cuts yield fifty-eight systems. 

Circumbinary planets are omitted because the rotation periods of circumbinary planets are not constrained by observation \citep{Banik2025}, and $\Omega$ enters $K$ directly through $\tau_{\rm eff}$. The sample thus contains fifty-eight eccentric planets, five gravity-darkened hosts and three pulsating hosts.

\bsp	
\label{lastpage}

\end{document}